\documentclass[authoryear,article,12pt]{elsarticle}

\makeatletter
\def\ps@pprintTitle{%
  \let\@oddhead\@empty
  \let\@evenhead\@empty
  \let\@oddfoot\@empty
  \let\@evenfoot\@oddfoot
}
\makeatother

\RequirePackage{tikz} 

\RequirePackage{amssymb}
\RequirePackage{amsthm}
\RequirePackage{enumerate}
\RequirePackage{color}
\RequirePackage{placeins}
\RequirePackage{pdfsync}
\RequirePackage{mathbbol}
\RequirePackage{natbib}
\RequirePackage{amsmath}
\RequirePackage{amsthm}
\RequirePackage[colorlinks]{hyperref}
\RequirePackage{xcolor}
\RequirePackage{array}
\RequirePackage{makecell}
\RequirePackage{booktabs}
\RequirePackage{fullpage}
\RequirePackage{graphicx} 
\RequirePackage{float} 
\RequirePackage{subcaption}
\RequirePackage{comment}
\RequirePackage{algorithm}
\usepackage{xr}
\usepackage{cleveref}
\RequirePackage{algpseudocode}
\definecolor{halfgray}{gray}{0.55} 
\definecolor{webgreen}{rgb}{0,.5,0}
\definecolor{webbrown}{rgb}{.6,0,0}
\definecolor{RoyalBlue}{rgb}{0,0.08,0.45}
\RequirePackage{hypernat}
\RequirePackage{multirow}
\RequirePackage{textcomp}
\usepackage{ulem}
\makeatletter
\NewDocumentCommand{\MakeTitleInner}{ +m +m +m }{
    \newpage%
    \null%
    \vskip 2em%
    \begin{center}%
        \let \footnote \thanks
        {\LARGE #1 \par}
        \vskip 1.5em%
        {%
            \large
            \lineskip .5em%
            \begin{tabular}[t]{c}%
                #2
            \end{tabular}\par%
        }%
        \vskip 1em%
        {\large #3}
    \end{center}%
    \par
    \vskip 1.5em%
}
\NewDocumentCommand{\MakeTitle}{ +m +m +m }{%
    \begingroup
        \renewcommand\thefootnote{\@fnsymbol\c@footnote}%
        \def\@makefnmark{\rlap{\@textsuperscript{\normalfont\@thefnmark}}}%
        \long\def\@makefntext##1{\parindent 1em\noindent
            \hb@xt@1.8em{%
                \hss\@textsuperscript{\normalfont\@thefnmark}%
            }##1%
        }%
        \if@twocolumn
            \ifnum \col@number=\@ne
                \MakeTitleInner{#1}{#2}{#3}
            \else
                \twocolumn[\MakeTitleInner{#1}{#2}{#3}]%
            \fi
        \else
            \newpage
            \global\@topnum\z@   
            \MakeTitleInner{#1}{#2}{#3}
        \fi
        \thispagestyle{plain}\@thanks
    \endgroup
    \setcounter{footnote}{0}%
    \setcounter{section}{0}%
    \setcounter{figure}{0}%
    \setcounter{table}{0}%
}
\makeatother

\usepackage{etoolbox}

\makeatletter
\newcommand*{\sortentry}[1]{%
  \if@filesw
    \immediate\write\@auxout{\string\uNAT@aux@sortentry{#1}}%
  \fi}
\newcommand*{\uNAT@aux@sortentry}{%
  \listgadd{\uNAT@bibsortlist}}
\newcommand*{\uNAT@bibsortlist}{}

\newcommand*{\uNAT@citekeys}{}

\newcommand*{\uNAT@writetocitelistsort}[1]{%
  \ifinlist{#1}{\uNAT@citekeys}
    {\ifdefvoid{\NAT@cite@list}
       {\def\NAT@cite@list{#1}}
       {\expandafter\def\expandafter\NAT@cite@list\expandafter{\NAT@cite@list,#1}}%
     \listgadd{\uNAT@foundkeys}{#1}}
    {}}

\newcommand*{\uNAT@writetocitelistforgotten}[1]{%
  \ifinlist{#1}{\uNAT@foundkeys}
    {}
    {\ifdefvoid{\NAT@cite@list}
       {\def\NAT@cite@list{#1}}
       {\expandafter\def\expandafter\NAT@cite@list\expandafter{\NAT@cite@list,#1}}}}

\newcommand*{\uNAT@sortcites}[1]{%
  \let\NAT@cite@list\@empty
  \let\uNAT@citekeys\@empty
  \let\uNAT@foundkeys\@empty
  \forcsvlist{\listadd{\uNAT@citekeys}}{#1}%
  \forlistloop{\uNAT@writetocitelistsort}{\uNAT@bibsortlist}%
  \forlistloop{\uNAT@writetocitelistforgotten}{\uNAT@citekeys}%
}

\def\NAT@citex%
  [#1][#2]#3{%
  \NAT@reset@parser
  \NAT@sort@cites{#3}%
  \uNAT@sortcites{#3}
  \NAT@reset@citea
  \@cite{\let\NAT@nm\@empty\let\NAT@year\@empty
    \@for\@citeb:=\NAT@cite@list\do
    {\@safe@activestrue
     \edef\@citeb{\expandafter\@firstofone\@citeb\@empty}%
     \@safe@activesfalse
     \@ifundefined{b@\@citeb\@extra@b@citeb}{\@citea%
       {\reset@font\bfseries ?}\NAT@citeundefined
                 \PackageWarning{natbib}%
       {Citation `\@citeb' on page \thepage \space undefined}\def\NAT@date{}}%
     {\let\NAT@last@nm=\NAT@nm\let\NAT@last@yr=\NAT@year
      \NAT@parse{\@citeb}%
      \ifNAT@longnames\@ifundefined{bv@\@citeb\@extra@b@citeb}{%
        \let\NAT@name=\NAT@all@names
        \global\@namedef{bv@\@citeb\@extra@b@citeb}{}}{}%
      \fi
     \ifNAT@full\let\NAT@nm\NAT@all@names\else
       \let\NAT@nm\NAT@name\fi
     \ifNAT@swa\ifcase\NAT@ctype
       \if\relax\NAT@date\relax
         \@citea\NAT@hyper@{\NAT@nmfmt{\NAT@nm}\NAT@date}%
       \else
         \ifx\NAT@last@nm\NAT@nm\NAT@yrsep
            \ifx\NAT@last@yr\NAT@year
              \def\NAT@temp{{?}}%
              \ifx\NAT@temp\NAT@exlab\PackageWarningNoLine{natbib}%
               {Multiple citation on page \thepage: same authors and
               year\MessageBreak without distinguishing extra
               letter,\MessageBreak appears as question mark}\fi
              \NAT@hyper@{\NAT@exlab}%
            \else\unskip\NAT@spacechar
              \NAT@hyper@{\NAT@date}%
            \fi
         \else
           \@citea\NAT@hyper@{%
             \NAT@nmfmt{\NAT@nm}%
             \hyper@natlinkbreak{%
               \NAT@aysep\NAT@spacechar}{\@citeb\@extra@b@citeb
             }%
             \NAT@date
           }%
         \fi
       \fi
     \or\@citea\NAT@hyper@{\NAT@nmfmt{\NAT@nm}}%
     \or\@citea\NAT@hyper@{\NAT@date}%
     \or\@citea\NAT@hyper@{\NAT@alias}%
     \fi \NAT@def@citea
     \else
       \ifcase\NAT@ctype
        \if\relax\NAT@date\relax
          \@citea\NAT@hyper@{\NAT@nmfmt{\NAT@nm}}%
        \else
         \ifx\NAT@last@nm\NAT@nm\NAT@yrsep
            \ifx\NAT@last@yr\NAT@year
              \def\NAT@temp{{?}}%
              \ifx\NAT@temp\NAT@exlab\PackageWarningNoLine{natbib}%
               {Multiple citation on page \thepage: same authors and
               year\MessageBreak without distinguishing extra
               letter,\MessageBreak appears as question mark}\fi
              \NAT@hyper@{\NAT@exlab}%
            \else
              \unskip\NAT@spacechar
              \NAT@hyper@{\NAT@date}%
            \fi
         \else
           \@citea\NAT@hyper@{%
             \NAT@nmfmt{\NAT@nm}%
             \hyper@natlinkbreak{\NAT@spacechar\NAT@@open\if*#1*\else#1\NAT@spacechar\fi}%
               {\@citeb\@extra@b@citeb}%
             \NAT@date
           }%
         \fi
        \fi
       \or\@citea\NAT@hyper@{\NAT@nmfmt{\NAT@nm}}%
       \or\@citea\NAT@hyper@{\NAT@date}%
       \or\@citea\NAT@hyper@{\NAT@alias}%
       \fi
       \if\relax\NAT@date\relax
         \NAT@def@citea
       \else
         \NAT@def@citea@close
       \fi
     \fi
     }}\ifNAT@swa\else\if*#2*\else\NAT@cmt#2\fi
     \if\relax\NAT@date\relax\else\NAT@@close\fi\fi}{#1}{#2}}
\makeatother

\theoremstyle{remark}

\newcommand{\E}{\ensuremath{\mathbb{E}}}
\renewcommand{\Pr}{\ensuremath{\mathbb{P}}}

\renewcommand{\hat}{\widehat}
\renewcommand{\tilde}{\widetilde}

\newcommand{\vague}{\stackrel{\lower0.2ex\hbox{$\scriptscriptstyle
                    \it{v} $}}{\to}}
\newcommand{\weak}{\stackrel{\lower0.2ex\hbox{$\scriptscriptstyle
                    \it{w} $}}{\to}}
\newcommand{\what}{\stackrel{\lower0.2ex\hbox{$\scriptscriptstyle
                    \it{\hat{w}} $}}{\to}}
\newcommand{\eqdis}{\stackrel{\lower0.2ex\hbox{$\scriptscriptstyle
                    \mathrm{d}$}}{=}}
\newcommand{\distr}{\stackrel{\lower0.2ex\hbox{$\scriptscriptstyle
                    \it{d} $}}{\to}}
\begin{document}

\begin{frontmatter}



\title{Flexible space-time models for extreme data}


\author[inst1]{Lorenzo Dell'Oro}

\affiliation[inst1]{organization={Dipartimento di Scienze Statistiche, Università di Padova},
            city={Padova},
            country={Italy}}

\author[inst2]{Carlo Gaetan}

\affiliation[inst2]{organization={Dipartimento di Scienze Ambientali, Informatica e Statistica, Università Ca' Foscari di Venezia},
            city={Venezia},
            country={Italy}}

\begin{abstract}
Extreme value analysis is an essential methodology in the study of rare and extreme events, which hold significant interest in various fields, particularly in the context of environmental sciences. 
Models that employ the exceedances of values above suitably selected high thresholds possess the advantage of capturing the ``sub-asymptotic'' dependence of data.
This paper presents an extension of  spatial random scale mixture models to the spatio-temporal domain. A comprehensive framework for characterizing the dependence structure of extreme events across both dimensions is provided. Indeed, the model is capable of distinguishing between asymptotic dependence and independence, both in space and time, through the use of parametric inference.
The high complexity of the likelihood function for the proposed model necessitates a simulation approach based on neural networks for parameter estimation, which leverages summaries of the sub-asymptotic dependence present in the data. The effectiveness of the model in assessing the limiting dependence structure of spatio-temporal processes is demonstrated through both simulation studies and an application to rainfall datasets.
\end{abstract}




\end{frontmatter}


\section{Introduction}
The analysis of spatial extreme data has received a significant boost because many natural extreme hazards, such as heat waves, heavy rain and snowfall, high tides, and windstorms, have a spatial extent.
A significant portion of the research has focused on the development of spatial models grounded in the theory of extreme values, including max-stable processes \citep{de_Haan:1984,Smith:1990,Schlather:2002}. These processes emerge as the pointwise maxima taken over an infinite number of independent copies of appropriately rescaled stochastic processes.
In addition to the solid theoretical arguments, the widespread use of these models is also due to the possibility of estimating their parameters in a relatively simple way, at least for a not large number of sites \citep{Padoan:Ribatet:Sisson:2012, Castruccio:Huser:Genton:2016}.

Other spatial statistical models for extreme data \citep{Davison:Padoan:Ribatet:2012, Huser:Wadsworth:2020} have roots in hierarchical models \citep{Casson:Coles:1999, Cooley:Nychka:Naveau:2007, Gaetan:Grigoletto:2007,Sang:Gelfand:2009} and copula models
\citep{Bortot:Coles:Tawn:2000,Sang:Gelfand:2010}.

In a recent paper \citet{huser2024modelingspatialextremesenvironmental}  argue that although the max-stable processes are supported by a well-established asymptotic theory, their application in environmental studies exhibits some restrictions.
Since max-stable processes are defined in terms of block maxima at each location, they are not intended to describe the stochastic variability of the extremes of the original individual events.
Moreover, max-stable processes attempt to characterize the extremal dependence of the normalized maxima as the block size grows to infinity. In such a situation, two cases arise: dependence in the limit, i.e., asymptotic dependence, or exact independence, i.e., for any block size. 
However, what is often observed in reality is a third situation, in which the dependence vanishes in the limit (asymptotic independence), but is still present for finite block size (sub-asymptotic dependence).
For this reason, several attempts are made to build sub-asymptotic models for spatial extremes based on the exceedances of high thresholds, combining tail flexibility with computational tractability, see for example
\citet{Wadsworth:Tawn:2012,Bacro:Gaetan:Toulemonde:2016,Huser:Opitz:Thibaud:2017,Huser:Wadsworth:2019,Wadsworth:Tawn:2022}.

If the interest is to model the extremes of the original events, a further complication arises when dealing with spatio-temporal data. 
Many spatial models are fitted by ignoring the temporal dependence of the extremes and then adjusting the uncertainty of the parameter estimates \citep{Fawcett:Walshaw:2006}.
Modeling the spatio-temporal extremal dependence adds an extra layer of difficulties \citep{Tawn:Shooter:Towe:Lamb:2018, Huser:Davison:2014,Steinkohl:Davis:Kluppelberg:2013,Morris.al.2017, Bacro:Gaetan:Opitz:Toulemonde:2020}.
In fact, different forms of asymptotic dependence could arise in the spatial and temporal domains. The limiting class could itself change with the temporal and/or spatial lag.
In the aforementioned references, the models maintain the same limiting dependence in both space and time.

Recently, \citet{Simpson:Wadsworth:2021}, \citet{Simpson:Opitz:Watsworth:2023} provide examples of formulations for threshold exceedances that allow for different forms of extremal dependence in the two domains.
These papers use a conditional approach based on an asymptotic approximation of the distribution of the space-time process conditional on the value at a single location and time being extreme. The construction is highly flexible; however, it is unclear whether a unique space-time process exists, given that the specification is conditional.

Similarly, \citet{bortot2024model} employ a time series approach, wherein extreme spatial dependence is integrated into a time series model \citep{Davis:Mikosch:2008}. One of the advantages of the proposed model is that it allows for straightforward simulation, thereby facilitating the extrapolation of extremal functionals of interest. 
Nevertheless, this approach is limited in that the specific type of extremal dependence in the two domains must be predetermined, and the space-time interactions lack sufficient flexibility.

\cite{Huser:Wadsworth:2019}   introduced a scale mixture model that considers many different possibilities within the spatial framework. These possibilities range from the limit of asymptotic independence to the limit of complete dependence.
Subsequently, the spatial model was modified by 
\citet{Zhang:Shaby:Wadsworth:2022} for higher-dimensional inference, and extended by 
\citet{Majumder:Reich:Shaby:2024} to enable enhanced flexibility in the spatial dependence, and by \citet{Majumder:Reich:2023} to accommodate dynamic spatial dependencies over time.

In this paper, we propose ways to extend the scale mixture model to include spatial and temporal dimensions, which would allow greater flexibility than the space-time models previously considered in \citet{bortot2024model}, since the models presented here include both extremal dependence classes according to the value of a parameter that can be identified from the data.

The proposed models are employed as copula models and they are subsequently applied to a real data set. However, likelihood-based inference for the parameters of the models becomes infeasible even for moderate sample sizes. 

As the simulation of data from the proposed models is a relatively straightforward and rapid process, the parameter estimation is founded on a simulation-based methodology which employs neural networks to learn the implicit mapping between the simulated data and the parameter values \citep{gerber2021fast,Lenzi:Bessac:Rudi:Stein:2023,sainsbury2024likelihood,richards2023likelihood}.
Other recent works that make use of neural networks to identify the extremal dependence of spatial data include \cite{ahmed2022recognizing} and \cite{wixson2024neural}.

The remainder of the paper is organized as follows: in Section \ref{section:model}, we introduce our model. Section \ref{section:estimation_method} presents and discusses parameter estimation using neural networks. The estimation procedure is evaluated in Section \ref{section:application} through a small simulation study and an analysis of precipitation data from a dataset of rainfall amounts recorded at weather stations in the North Brabant province of the Netherlands. The paper concludes in Section \ref{section:conclusions}.

\section{A model for space-time extreme values}\label{section:model}
We will consider $Y(s,t)$   a spatio-temporal process, such that $s\in S \subset \mathbb{R}^d$ indicates a spatial location  on the geographic space $S$ and $t\in \mathcal{T}= \{1, 2,\ldots\,\}$ a discrete time.
To model the spatio-temporal extremal dependence of $Y(s,t)$, we follow a copula-based approach: first, we define a process, $X(s,t)$, with a suitable extremal dependence structure, and then we marginally transform it into $Y(s,t)$. 
Moreover, since the goal is to model only extreme values, we set a high threshold and consider the values that exceed this threshold, while censoring the other observations. In this section, we focus on the specification of $X(s,t)$, leaving the details of marginal transformation to the data application (see Section \ref{section:application}).

We begin by recalling some background material on extremal dependence.

\subsection{Asymptotic dependence and independence classes}

Let  $X_{1}:=X(s_1,t_1)$ and $X_{2}:=X(s_2,t_2)$ be a generic pair of random variables of a space-time process $X(s,t)$ with marginal distribution functions $F_{X_1}$ and $F_{X_2}$, respectively.
The strength of dependence in the upper tail of $(X_1,X_2)$ is commonly quantified in terms of the tail dependence coefficient, $\chi \in [0, 1]$, defined as the limit  of
$$
\chi(u)=\frac{\Pr\left(F_{X_1}(X_{1})>u,F_{X_2}(X_{2})>u \right)}{1-u}
$$
when this limit exists \citep{Joe:1997}, i.e.
\begin{equation}\label{def:chi_X_12}
	\chi=\lim_{u\rightarrow 1^-}\chi(u)
	=\lim_{x\rightarrow\infty}\frac{\Pr\left(X_{1}>x,X_{2}>x\right)}{\Pr\left(X_{1}>x\right)}.
\end{equation}
The last equivalence holds if we assume the marginal distribution of $X(s,t)$ to be stationary over $\mathcal{S}$ and $\mathcal{T}$, with the upper endpoint at infinity.

When $\chi > 0$, the components of $(X_1,X_2)$ are said to be asymptotically dependent (AD) in the upper tail, and when $\chi = 0$, they are said to be asymptotically independent (AI).
 
In the latter case it is useful to  consider the rate at which the subasymptotic measure $\chi(u)$ tends to zero as $u\rightarrow 1^-$. Following \citet{Ledford:Tawn:1996} we may assume that
\begin{equation}\label{eq:LTcondition}
\chi(u)\sim \ell\left((1-u)^{-1}\right) (1-u)^{1/\eta-1},
\end{equation}
where $\ell(\cdot)$ is a slowly varying function at infinity, that is, $\ell(tx)/\ell(x) \rightarrow 1$  as $x\rightarrow\infty$ for all $t > 0$.
The coefficient $0<\eta\le 1$ is called the residual tail dependence coefficient. 
When $\eta = 1$ and $\lim_{u\rightarrow 1^-}\ell((1-u)^{-1})\ne 0$ we obtain AD, but otherwise there is AI.
The case $\eta <1$  may be further classified into (a) positive association with $0.5<\eta <1$; (b) near-independence with  $\eta=0.5$; and (c) negative association with $0<\eta <0.5$.
Here, for notation simplicity,  we have omitted any reference to the pair $(X_1, X_2)$ when defining $\chi(u)$, $\chi$, $\eta$ and $\ell$.

Finally, we say that the process $X(s,t)$ is asymptotically independent (AI) or asymptotically dependent (AD) if $\eta$ is either strictly less than $1$ or equal to $1$, respectively, for all pairs $(X_1,X_2)$. 
Moreover  we say that the process $X(s,t)$ is AI or AD in space if  $\eta$ is either strictly less than $1$ or equal to $1$, respectively, for all pairs $(X(t,s_1),X(t,s_2))$.
Similar definition can be stated in case of the time.

\subsection{Model definition}

Our starting point is the spatial scale mixture model proposed by 
\citet{Huser:Wadsworth:2019}, namely 
\begin{equation}\label{model_HW}
X(s)=R^{\delta}\, W(s)^{1-\delta},\quad {\delta}\in[0,1],
\end{equation}
where $R$ is a standard Pareto random variable, i.e.
	$\Pr(R\le r)=1-r^{-1}$, $r\ge 1$,
and $W(s)$ is an AI spatial process, independent of $R$, whose marginal distribution is also standard Pareto. 
The strength of this specification is that it can cover the AD and AI case according the value of $\delta$. More precisely, 
\citet{Huser:Wadsworth:2019} show that if ${\delta}>0.5$ then $X(s)$ is AD, while if ${\delta}\le0.5$, $X(s)$ is AI.

A direct extension of the model \eqref{model_HW} for space-time data would be $X(s,t)=R^{\delta}\, W(s,t)^{1-\delta}$, ${\delta}\in[0,1]$, where $W(s,t)$ is a space-time process. This model would be characterized by either asymptotic dependence in both space and time, or asymptotic independence in both space and time, depending on the value of $\delta$. Therefore, it may be too rigid since it is not able to cover mixed extremal dependence classes in the two dimensions.

We propose the following space-time model:
\begin{equation}\label{model_X_RW}
X(s,t)=R(t)^\delta \, W(s,t)^{1-\delta}, \quad \delta\in [0,1],
\end{equation}
where the processes $R(t)$ and $W(s,t)$ are mutually independent with standard Pareto marginal distribution.
Making $R$ depend on $t$ provides more flexibility to the model, allowing to get different extremal dependence classes in space and time, as shown in Section \ref{section:dependence_properties}.
The marginal distribution of $X(s,t)$  is given by 
\begin{equation}\label{marginal_distr_X}
 G(x)= \left\{
 \begin{array}{ll}
  1-\left[\delta/(2\delta-1)\, x^{-1/\delta}-(1-\delta)/(2\delta-1)\, x^{-1/(1-\delta)}\right] & \text{for }\delta\neq 0.5,\\
  1-x^{-2}[2\log(x)+1] &\text{for } \delta=0.5.
 \end{array} 
 \right.
\end{equation}
An alternative formulation of model \eqref{model_X_RW} can be obtained by taking the logarithm
\begin{equation}\label{model_log_X_RW}
\tilde{X}(s,t)=\delta\tilde{R}(t) + (1-\delta) \tilde{W}(s,t),
\end{equation}
where $\tilde{X}(s,t):=\log\{X(s,t)\}$, $\tilde{R}(t) := \log\{R(t)\}$ and $\tilde{W}(s,t):=\log\{W(s,t)\}$. In such case the marginal distributions of 
$\tilde{R}(t)$ and $ \tilde{W}(s,t)$ are standard exponential  distributions and the marginal distribution of $\tilde{X}(s,t)$ is known as hypo-exponential or general Erlang distribution \citep[p.\ 552]{johnson1994continuous}.
The formulation \eqref{model_log_X_RW} preserves the extremal dependence structure of \eqref{model_X_RW}, but is computationally more stable, making it preferable for simulation and parameter estimation.

The representation \eqref{model_log_X_RW} shares some similarities  with the model presented in \citet{Majumder:Reich:2023}, namely 
\begin{equation*}\tilde{X}_t(s)={\delta(t)}\,\tilde{R}_t(s)+ (1-\delta(t))\,\tilde{W}_t(s),\quad {\delta(t)}\in[0,1],
\end{equation*}
where $\tilde{R}_t(s)$, $t=1,\ldots,$ are independent copies of an AD spatial process, and $\tilde{W}_t(s)$, $t=1,\ldots,$ are independent copies of an AI spatial process.
However, the authors of the aforementioned paper primarily focused on the spatial extremal dependence. They allowed for the possibility that this spatial dependence may not be stationary in time by varying the delta function. 
In contrast, the present study posits that the extremal dependence is driven by two processes, denoted $\tilde{R}(t)$ and $\tilde{W}(s,t)$, which exhibit time and space-time dependence.

Before illustrating the extremal dependence properties of the model \eqref{model_X_RW}, we highlight its inherent asymmetry. The current definition uses two distinct stochastic processes to model temporal dependence and one to model spatial dependence. An alternative is \begin{equation}\label{eq:model_5678}
    X(s,t)=R(s)^\delta\, W(s,t)^{1-\delta},
\end{equation}
where more flexibility is allocated to the spatial dimension. 
Our selection of \eqref{model_X_RW} is motivated by its application in our real-data analysis. However, studying the extremal dependence properties of the alternative model would not introduce additional complexity if we carefully invert the roles of the spatial and temporal dimensions. Table \ref{table:possible_models_5678} in the Supplementary material summarizes the dependence properties of models derived from equation \eqref{eq:model_5678}.

\subsection{Dependence properties of the model}\label{section:dependence_properties}

In principle, both $R(t)$ and $W(s,t)$ in \eqref{model_X_RW} can be AI or AD stochastic processes. 
This results in four different models for $X(s,t)$ with different extremal dependence, which are summarized in Table \ref{table:possible_models}. 
The Table shows the extremal dependence for pairs $X(s_1,t_1),X(s_2,t_2)$ in space (i.e.,  for $s_1\neq s_2,\;t_1=t_2$), in time (i.e., for $s_1=s_2,\;t_1\neq t_2$) and in space-time (i.e., for $s_1\neq s_2,\;t_1\neq t_2$), for different values of the parameter $\delta$.

The reader is referred to the appendices for a detailed exposition of the proofs in the space-time case. These proofs adopt a similar approach to that of 
\cite{engelke2019extremal}. In particular, \ref{section:appendix_time_spacetime} contains the proofs of the results for pairs in time and space-time, while \ref{section:appendix_space} refers to pairs in space.

\begin{table}[ht!]
	\begin{center}
		\begin{tabular}{| c | c | c | c | c | c |} 
			\hline
			Model & $R(t)$ & $ W(s,t)$ & $\delta>0.5$ & $\delta=0.5$ & $\delta<0.5$ \\ [0.5ex] 
			\hline\hline
			1 & AI & AD & \makecell[l]{AD in space\\ AI in time\\AI in space-time} & \makecell[l]{AD in space\\ AI in time\\AI in space-time} & \makecell[l]{AD in space\\ AD in time\\AD in space-time} \\ 
			\hline
			2 & AD & AI & \makecell[l]{AD in space\\ AD in time\\AD in space-time} & \makecell[l]{AI in space\\ AI in time\\AI in space-time} & \makecell[l]{AI in space\\ AI in time\\AI in space-time} \\ 
			\hline
			3 & AI & AI & \makecell[l]{AD in space\\ AI in time\\AI in space-time} & \makecell[l]{AI in space\\ AI in time\\AI in space-time} & \makecell[l]{AI in space\\ AI in time\\AI in space-time} \\ 
			\hline
			4 & AD & AD & \makecell[l]{AD in space\\ AD in time\\AD in space-time} & \makecell[l]{AD in space\\ AD in time\\AD in space-time} & \makecell[l]{AD in space\\ AD in time\\AD in space-time} \\ 
			\hline
		\end{tabular}
		\caption{The four possible combinations for  $R(t)$ and $W(s,t)$ and the resulting extremal dependence for pairs $[X(s_1,t_1), X(s_2,t_2)]$ in space, in time and in space-time, for different values of the parameter $\delta$. These statements are supported by proof  in \ref{section:appendix_time_spacetime} and \ref{section:appendix_space}.}
		\label{table:possible_models}
	\end{center}
\end{table}
With the exception of Model 4, all other models are characterized by two different extremal dependence configurations in space and time. The selection between these configurations is dependent on the value of the parameter $\delta$. 
In particular, $\delta$ controls which of the two components, $R(t)$ or $W(s,t)$, dominates in the mixture and leads the extremal dependence. 
This allows for the comparison and testing of a pair of scenarios specific to each model. 
When $\delta>0.5$, $X(s,t)$ belongs to the same extremal dependence class as $R(t)$ for spatio-temporal pairs and also for temporal pairs. However, for $t=\bar{t}$ fixed, $R(\bar{t})$ is a single random variable, so it is exactly dependent for spatial pairs, regardless of the dependence class of $R(t)$; then, these pairs are always AD when $\delta>0.5$. On the other hand, when $\delta<0.5$, $W(s,t)$ dominates the mixture and controls the extremal dependence for all the pairs of $X(s,t)$. When $\delta=0.5$, the pairs are AI if they are also AI for at least one value of $\delta\neq 0.5$, otherwise they are AD.

This subsection concludes by introducing two specifications of processes, $R(t)$ and $W(s,t)$, that we use in our numerical examples.
They are derived by transforming two other underlying processes, denoted by $R^*(t)$ and $W^*(s,t)$.

The most prevalent and extensively utilized class of AI processes is that of Gaussian processes. Here we consider Gaussian processes with underlying correlation function $\rho(l_1,l_2)$ for two generic time or space-time locations $l_1$ and $l_2$.
The Gaussian processes satisfy \eqref{eq:LTcondition}
with $\eta(l_1,l_2) = \{1 + \rho(l_1,l_2)\}/2$ \citep{Ledford:Tawn:1996,Sibuya:1960}, so that $\eta(l_1 , l_2) < 1$ (asymptotic independence) whenever $\rho(l_1,l_2) < 1$.
Their usefulness as a model for $R(t) $ or $W(s,t) $ is motivated by the wealth of examples of parametric correlation functions and by the relative simplicity of the simulation.

On the other hand, max-stable processes are an important class of AD processes. However, their simulation and inference are computationally prohibitive in high dimensions \citep[see, for instance,][]{Huser:Wadsworth:2020}.

The class of Student's \textit{t} processes \citep{Roislien:Omre:2006} provides a manageable alternative. Student's \textit{t} processes with $\nu>0$ degrees of freedom are defined as zero mean and unit variance Gaussian processes, with correlation function $\rho(l_1,l_2)$, divided by the square root of a Gamma random variable, with shape parameter $\nu/2$ and rate parameter $\nu/2$. They are AD processes \citep{chan:li:2008} and converge to the Gaussian limit as $\nu \rightarrow \infty$.

Therefore, in the following we consider two processes, which can be Gaussian or Student's \textit{t} processes with $\nu$ degrees of freedom, named $R^*(t)$ and $W^*(s,t)$. If we assume that both processes are stationary, they  are identified by the correlation functions $\rho_{R^*}(k;\phi)$ and $\rho_{W^*}(h,k;\psi)$, where $k \in \mathbb{R}$ is a temporal lag and $h\in \mathbb{R}^2$ is a spatial lag, while $\phi$ and $\psi$ are two sets of parameters.
The transformations 
\begin{eqnarray}\label{eq:transform_pareto}
    R(t) &=& 1 / [1 - F_{R^*}(R^*(t))],\nonumber\\ 
    W(s,t) &=& 1 / [1 - F_{W^*}(W^*(s,t))]
\end{eqnarray}
where $F_{R^*}$ and $F_{W^*}$ are the marginal distribution functions of $R^*(t)$ and $W^*(s,t)$, cause $R(t)$ and $W(s,t)$ to marginally follow a standard Pareto distribution and preserves the extremal dependence class of the original processes.

\subsection{A model for the marginals}\label{section:model_marginal}

The model specified by \eqref{model_X_RW} is employed as a copula to describe the extremal dependence of the data $Y(s,t)$.
To model extreme values of the observation $Y(s,t)$, we employ a Peaks Over Thresholds (POT) approach \citep{Davison:Smith:1990,Eastoe:Tawn:2009}. This approach involves fixing a threshold, denoted by $\mu(s,t)$, which represents the $p$-quantile of the distribution of $Y(s,t)$ at location $s$ and time $t$, and using a generalized Pareto distribution as the stochastic model for the exceedances over the threshold.

More precisely, it is assumed that $X(s,t)$, as defined in \eqref{model_X_RW}, is a stationary process and $G$ is its marginal distribution, as defined in \eqref{marginal_distr_X}. Then, the values of the process $X(s,t)$ are transformed into 
\begin{equation}\label{eq:marginal_Y}
    Y(s,t) = Q_{s,t}(G(X(s,t))),
\end{equation}
where
\begin{eqnarray}\label{eq:F_inv_GPD}
    Q_{s,t}(u) &=& 
    \left\{
    \begin{array}{ll}
    \mu(s,t) & \mbox{for }  0<u\le p,\\
    \mu(s,t) +\sigma \left[\left\{1-(u-p)/(1-p)\right\}^{-\xi} -1 \right]/\xi & \mbox{for } p<u< 1,\; \xi\ne0,\\
    \mu(s,t) -\sigma \log\left[1-(u-p)/(1-p)\right] & \mbox{for } p<u< 1,\; \xi=0,
    \end{array}
    \right.
\end{eqnarray}
with $0<p<1$.
In the data application (Section \ref{section:application}), $\mu(s,t)$ is preliminary estimated via a spatially dependent quantile regression, while the estimation method for the other parameters is detailed in Section \ref{section:estimation_method}.
Note that, although $\sigma$ and $\xi$ could also theoretically depend on space or time, we assume here that they are constant based on the real data example.
Also note that the truncation operated in \eqref{eq:F_inv_GPD} does not affect the values of the coefficient $\chi$ in \eqref{def:chi_X_12} and of its sub-asymptotic version $\chi(u)$ with $u>p$.

\section{Estimation method}\label{section:estimation_method}
Our final stochastic model for the extreme values of $Y(s,t)$ is the result of composing the copula model \eqref{eq:transform_pareto} and \eqref{model_X_RW} and the marginal model \eqref{eq:marginal_Y}.
In the following, the set of parameters of the copula model is denoted as $\theta_D:=(\phi,\psi,\delta)$ and the set of marginal parameters is given by $\theta_M=(\sigma,\xi)$, assuming that the threshold $\mu(s,t)$ has been previously identified.

Estimation of the parameters of scale mixture models is typically accomplished through two sequential steps \citep{Huser:Opitz:Thibaud:2017,Huser:Wadsworth:2019}.
We will briefly outline these steps to highlight the complexity of adopting it in our case.

\medskip

1. The first step deals with  the marginal parameters $\theta_M$ of the cumulative distribution function (CDF) derived from \eqref{eq:marginal_Y} 
$$
 F_{s,t}(y) = \left\{\begin{array}{ll}
	 p  &\quad \text{for}\; y\le\mu(s,t),\\
	  p + (1-p) \left[1-\left(1+\xi(y-\mu(s,t))/\sigma\right)_+^{-1/\xi}\right] &\quad \text{for}\; y>\mu(s,t),\,\xi\ne0,\\
	p + (1-p) \left[1-\exp\{-(y-\mu(s,t))/\sigma\}\right] &\quad \text{for}\; y>\mu(s,t),\; \xi=0,
	   \end{array} \right.
$$
where $(a)_+ = \max(0, a)$. The estimation of $\theta_M$ can be based on the maximization of the independence likelihood \citep{Varin:Reid:Firth:2011}. This approach is used in our numerical examples.

\medskip

2. In the second step, it is assumed that for the observations above $\mu(s,t)$ the process $Y(s,t)$ has the same copula of $X(s,t)$.  
A likelihood-based inference approach for estimating $\theta_D$ requires evaluating the joint distribution of $X(s_i,t_j)$, $i=1,\ldots,n$, $j=1,\dots,T$, where $n$ is the number of observed spatial locations and $T$ is the number of observed time periods.

With $X_{i,j}=X(s_i,t_j)$, $R_j={R}(t_j)$, ${W}_{i,j}={W}(s_i,t_j)$ we can write the CDF of the vector ${\mathbf{X}}=({X}_{1,1},\dots,{X}_{1,T},{X}_{2,1},\dots,X_{2,T},\dots,X_{n,1},\dots,X_{n,T})$ as
\begin{eqnarray*}	
    F_{{\mathbf{X}}}(x_{1,1},\dots,x_{n,T}) &=&\Pr\left({X}_{1,1}\le x_{1,1},\dots,{X}_{n,T}\le x_{n,T}\right)\\
    &=&\Pr\left({R}_1^\delta{W}_{1,1}^{1-\delta}\le x_{1,1},\ldots,{R}_T^\delta{W}_{n,T}^{1-\delta}\le x_{n,T}\right)\\
    &=& \Pr\left({W}_{1,1}\le x_{1,1}^{1/(1-\delta)} {R}_1^{-\delta/(1-\delta)},\ldots,{W}_{n,T}\le x_{n,T}^{1/(1-\delta)} {R}_T^{-\delta/(1-\delta)}\right)\\
    &=& \int_{0}^{\infty}\cdots\int_{0}^{\infty}F_{{\mathbf{W}}}\left(x_{1,1}^{1/(1-\delta)} {r}_1^{-\delta/(1-\delta)},\ldots, x_{n,T}^{1/(1-\delta)} {r}_T^{-\delta/(1-\delta)}\right)\times\\
    && \qquad \qquad \quad f_{{\mathbf{R}}}(r_1,\dots,r_T) \, d r_1\cdots dr_T	
\end{eqnarray*}
and the density function of ${\mathbf{X}}$ as
\begin{eqnarray*}
    f_{{\mathbf{X}}}(x_{1,1},\dots,x_{n,T}) &=&
    \frac{1}{(1-\delta)^{nT}}
    \prod_{i=1}^{n}\prod_{j=1}^{T} \left[x_{i,j}^{\delta/(1-\delta)}\right] \times \\ && \int_{0}^{\infty}\cdots\int_{0}^{\infty}
    f_{{\mathbf{W}}}\left(x_{1,1}^{1/(1-\delta)} {r}_1^{-\delta/(1-\delta)},\ldots, x_{n,T}^{1/(1-\delta)} {r}_T^{-\delta/(1-\delta)}\right) \times\\
    &&\qquad\qquad
    \prod_{j=1}^{T} \left[r_j^{-\delta/(1-\delta)}\right] f_{{\mathbf{R}}}(r_1,\dots,r_T) \, dr_1\cdots dr_T,
\end{eqnarray*}
where $F_{{\mathbf{W}}}$ and $f_{{\mathbf{W}}}$ indicate the CDF and the density of ${\mathbf{W}}= ({W}_{1,1},\dots,W_{n,T})$ and $f_{{\mathbf{R}}}$ the density of ${\mathbf{R}}= ({R}_1,\dots,R_T)$.
In general, there is no analytical formula for $F_\mathbf{X}$ or $f_\mathbf{X}$ and the evaluation of the $T$-dimensional integral is therefore the core of the computational task.
Note that in the spatial case of the model \eqref{model_HW}, the previous formulas reduce to a one-dimensional integral \citep[see][Eq. (15)-(16)]{Huser:Wadsworth:2019}.
On the other hand, the computation of bivariate distribution functions for model \eqref{model_X_RW} involves at least a double integral, making parameter estimation still infeasible even when inference is based on pairwise likelihoods as in \citet{Huser:Davison:2014}.

\medskip

Notwithstanding the challenges associated with implementing a likelihood-based approach for estimating $\theta_D$, it is noteworthy that simulating a large number of spatio-temporal datasets from the copula model is comparatively easy and fast. In the following, we propose an estimation of the parameters of interest based on a simulated approach inspired by the Neural Bayes estimators \citep{sainsbury2024likelihood,richards2023likelihood}.

The idea is to choose, or randomly generate, a dense configuration of values covering the entire parameter space, or a reasonable subset of it.
For each parameter value, a space-time dataset is then simulated.
The implicit mapping that connects data and parameters is estimated by a neural network. The same network can then be used to predict parameter values from datasets that are similar, in terms of parameter space, to those used for its training.
The variability of 
these point estimates is typically evaluated through a bootstrap procedure.

A major difference, compared to the Neural Bayes Estimators, is that we use appropriately chosen summary statistics as input of the neural networks, instead of letting the network extract them from the datasets. A similar approach has been implemented by \citet{gerber2021fast} in the geostatistical context.
Although the optimality of the choice of statistics is not guaranteed, this leads to less memory usage and faster network training, allowing the space-time dimensionality to be considerably increased.
In particular, we employ the empirical pairwise coefficient $\hat{\chi}(u)$, which has been previously used as input of a neural network by \cite{ahmed2022recognizing}. 
Since in the data application (Section \ref{section:application}) we have observations over $N=20$ years and we assume independence between different years, we can compute the coefficient as an average over $i=1,\dots,N$. Note that, inside each independent year $i$, we have space- and time-dependent observations;
we denote them as $y_{1,i}$, $y_{2,i}$  for the generic pair $Y_1=Y(s_1,t_1)$, $Y_2=Y(s_2,t_2)$, where $s_1$ and $s_2$ refer to spatial locations and $t_1$ and $t_2$ refer to the day of the year. We calculate the empirical coefficient as
\begin{equation}\label{empirical_chi}
    \hat{\chi}_{1,2}(u)=\frac{\sum_{i=1}^N \mathbb{1}(y_{1,i}>\hat{F}_1^{-1}(u),y_{2,i}>\hat{F}_2^{-1}(u))}{N(1-u)},
\end{equation}
where $\hat{F}_1^{-1}(u)$ and $\hat{F}_2^{-1}(u)$ are the estimated quantiles of order $u$ for the locations $s_1$ and $s_2$, with $u>p$ in \eqref{eq:F_inv_GPD}.
Moreover, we compute the average of $\hat{\chi}_{1,2}(u)$ for all the pairs $(s_1,t_1)$ and $(s_2,t_2)$ lying at $m_1$ different ranges of spatial distances and $m_2$ temporal lags, resulting in a grid of $m_1\times m_2$ values in $[0,1]$. This is done for $u\in\{0.90, 0.95, 0.99\}$, leading to three different grids meant to cover the sub-asymptotic dynamics of the spatio-temporal dependence of the data; a finer sequence of $u$ values would result in empirical estimates of $\chi(u)$ too similar for adjacent values of the threshold level. On the other hand, when values of $u>0.99$ are considered, significant variability in the estimates of $\chi(u)$ is observed, given the sample size of our data application, presented in Section \ref{section:application}.
In that case, $m_1=m_2=8$, with the longest distance in the grids set at $34$ km, i.e. the half of the maximum distance of the observed locations, and the time lags that range from $0$ to $7$ days, since we assume the dependence to be negligible for higher distances or lags. In case of non-regularly observed time periods, the grids could be defined by treating them as continuous variables, as we do with the spatial locations.
Note that \eqref{empirical_chi} is the empirical version of \eqref{def:chi_X_12} for a finite threshold, even though the first is defined for the process $Y(s,t)$, in the scale of the observed data, while the second refers to the copula process $X(s,t)$. Indeed, the marginal transformation linking the two processes does not affect the values of this coefficient.

In the network training, a two-dimensional convolution \citep{Hugo:Pinaya:Vieira:Garcia-Dias:Mechelli} is applied to the $m_1\times m_2$ grids of $\hat{\chi}(u)$'s, treating each of the three grids as a channel, as if they were the three RGB colours intensity of an image \citep{lecun2015deep}. 
The result is then passed through few dense layers, resulting in an output vector corresponding to the values of the parameters.
The loss function to be minimized by the network is the mean absolute error.
In our setting, the validation set consists of $20\%$ of the input data.
The neural network is built and trained using the R package \texttt{keras3} \citep{keras3} an interface to Keras, a high-level neural networks API. The training takes about one minute on a standard laptop, using the  Root Mean Square Propagation (RMSprop) optimization algorithm and 40 epochs for data split into batches of size 128. For more details on the implementation of the convolutional layers and the estimation algorithm, see the Keras page \texttt{https://keras.io}.

\begin{algorithm}[t]
\caption{Neural estimation method for $\theta_D$}\label{alg:estimation_method}
\begin{algorithmic}
\State \textbf{Simulation}
\begin{enumerate}
    \item Generate many parameter values $\theta_D^*$ uniformly on the parameter space
    \item Simulate a dataset $y^*$ from the model for each parameter value $\theta^*$
    \item Summarize each dataset $y^*$ through some statistics $z^*$
\end{enumerate}
\State \textbf{Point estimation} 
\begin{enumerate}
\setcounter{enumi}{3}
    \item Learn the  map $g(\cdot)$ that connects $\theta_D^*$ to the statistics $z^*$ using neural networks
    \item Use the  map to predict a value of $\theta_D$ from the statistics computed on the observed dataset, $z^{o}$. The predicted value $\hat{\theta}_D=g(z^o)$ is the neural estimate
\end{enumerate}
\end{algorithmic}
\end{algorithm}

The amount of input data for training and testing a neural network is a crucial parameter and there is no general rule.  This amount would depend on the number of parameters, the architecture of the neural network, and the optimization algorithm used.
In this study, an empirical approach is adopted, wherein the model is trained on datasets of increasing size, and the validation performance is observed. For our specific numerical examples, see Section \ref{section:application}, 
we have seen that if we randomly generate $K=30\,000$ parameter vectors on $\Theta_D$, we get stable results.  
For each of these parameter values, a dataset is simulated with the same spatial locations and time periods as the observed one. Each dataset is then summarized by the statistics described above. 
This simulation takes about 320 minutes on a 2.3 GHz machine with 32 cores and 30 GB of memory. We have also experimented that increasing $K$ up to $70\,000$ does not change significantly the validation performance.

The estimation method described above is summarized in Algorithm \ref{alg:estimation_method}. An R code implementing it is available on request from the authors. 
Note that the employed statistics, i.e. the empirical $\hat{\chi}(u)$ coefficients, focus on the joint tail (above $0.90$-quantiles) of the data distribution, without giving the network any information on its bulk; in this way, we are operating an implicit censoring on the data, without the need to adapt the network to a censoring scheme, as in \cite{richards2023likelihood}.

The variability of the point estimates for both $\theta_D$ an $\theta_M$ is evaluated through a parametric bootstrap approach: once we get the neural estimate $\hat{\theta}_D$, we use it to simulate $B=400$ datasets from $X(s,t)$, and apply the same network to each dataset, getting estimates $\hat{\theta}_D^b$, for $b=1,\dots,B$; moreover, we transform each dataset to the $Y(s,t)$ scale through \eqref{eq:marginal_Y}, using the maximum (independence) likelihood estimate $\hat{\theta}_M$, and we evaluate $\hat{\theta}^b_M$, for $b=1,\dots,B$.
We can then compute bootstrap confidence intervals as percentile intervals based on the univariate distribution of the estimates for each parameter.

\section{Numerical examples}\label{section:application}

\subsection{Rainfall data}

\begin{figure}[h!]
    \centering
    \includegraphics[width=0.55\textwidth]{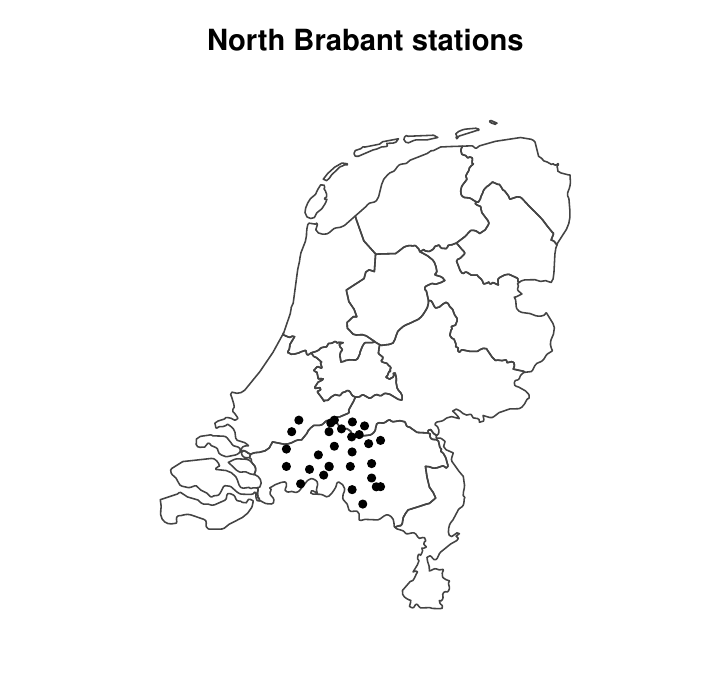}
    \caption{Location of the 30 stations in the North Brabant province of the Netherlands.}
    \label{fig:stations_NL}
\end{figure}

The models described in Section \ref{section:model} are applied to a dataset of daily rainfall collected between 1999 and 2018 in the North Brabant province of the Netherlands.
The dataset, already studied by \cite{bortot2024model}, was downloaded from the European Climate Assessment (ECA) and Dataset website (\href{https://www.ecad.eu}{https://www.ecad.eu}) and contains data collected over 30 stations, after discarding those with missing data.
Moreover, we focus here on spring months (March, April, May) to avoid seasonality. This leads to 20 blocks (each corresponding to one year) of 92 observations for each station. The data are assumed to be time-dependent within each year and independent between years.
Figure \ref{fig:stations_NL} shows the location of the 30 stations within the Netherlands. The relatively small size and the geographic homogeneity of the region support the hypothesis of strong spatial dependence of climate events (such as rainfall) within it, although this dependence is not necessarily persistent as the magnitude of the events becomes more extreme.

\begin{figure}[h!]
\begin{subfigure}[h]{0.5\linewidth}
\includegraphics[width=\linewidth]{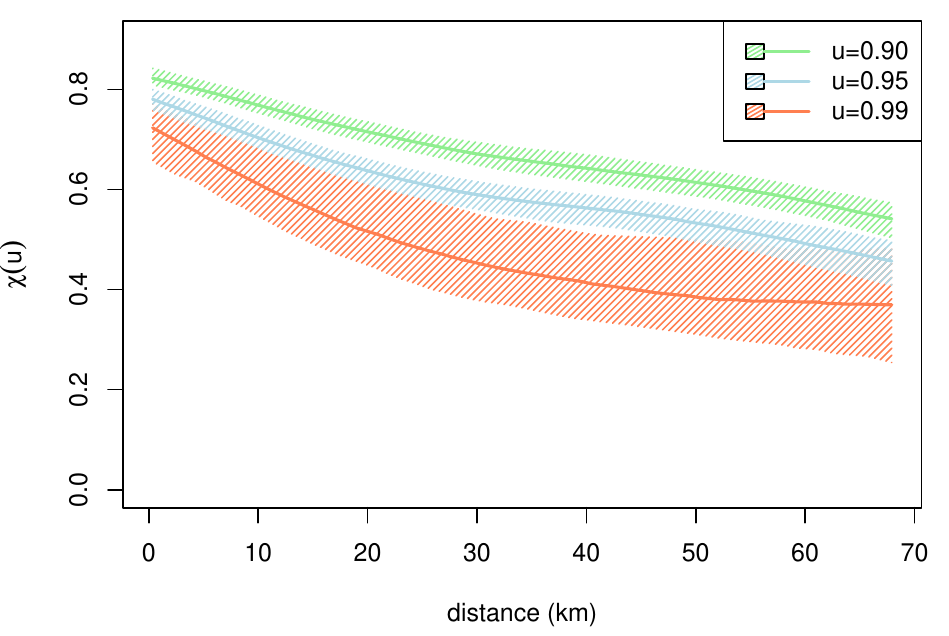}
\end{subfigure}
\hfill
\begin{subfigure}[h!]{0.5\linewidth}
\includegraphics[width=\linewidth]{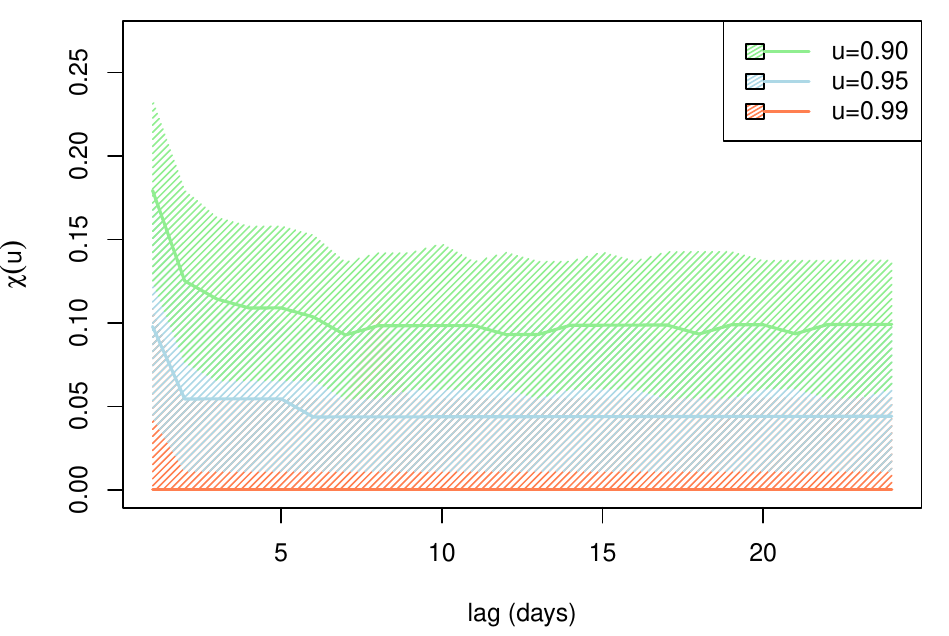}
\end{subfigure}%
\caption{For the North Brabant rainfall data, empirical estimates of $\chi(u)$, $u=0.90,0.95,0.99$, for pairs of observations at increasing spatial distances (left) and temporal lags (right). The shaded areas represent approximate $95\%$ confidence regions based on a stationary bootstrap procedure.}
\label{fig:chi_u_NL_space_time}
\end{figure}

Figure \ref{fig:chi_u_NL_space_time} displays the empirical estimates of the coefficient $\chi(u)$, defined in \eqref{empirical_chi}, on the North Brabant dataset, for $u\in\{0.90,0.95,0.99\}$, for pairs at different spatial distances, averaging across all time periods, and for pairs at different temporal lags, averaging across all sites. The shaded areas represent $95\%$ pointwise confidence intervals obtained by $200$ replications of a block bootstrap.
Although empirical estimates decrease with spatial and temporal distances, their limit as $u$ increases appears to be positive in the first case and zero in the second case. This may be an evidence of asymptotic dependence in space and asymptotic independence in time.

\subsection{Space-time model for rainfall data}\label{section:application_model}

\cite{bortot2024model} study the same data, although focusing on different months, and compare models with various space-time extremal dependence, concluding that the process underlying the rainfall data in the North Brabant province is AI in time and AD in space. This setting, which seems to be confirmed by exploratory analysis on our data (see Figure \ref{fig:chi_u_NL_space_time}), is possible only under Model 1 and Model 3 (see Table \ref{table:possible_models}), so we focus on these two configurations. Note that the value of $\delta$ has a different interpretation in the two models: Model 1 is always AD in space and could be AI or AD in time and space-time, depending on $\delta$; Model 3 is always AI in time and space-time and could be AI or AD in space, depending on $\delta$.

In Model 1 and Model 3, $R(t)$ is chosen to be asymptotically independent, more precisely, $R^*(t)$ is a Gaussian process with zero mean and unit variance, with correlation function $\rho_{R^*}(k;\phi)= \exp(-k/\phi)$, for temporal lag $k$.

The process $W(s,t)$ should be AD in Model 1 and AI in Model 3. In both cases, we again follow \eqref{eq:transform_pareto}, where $W^*(s,t)$ is a Gaussian or Student's \textit{t} process with $\nu=1$ degrees of freedom, respectively. The parameter $\nu$ could also be estimated, but we fix its value at 1 to limit the number of parameters. Moreover, as $\nu\rightarrow\infty$, the Student's \textit{t} process tends to a Gaussian process, i.e., an AI process.
Thus, setting a low value for $\nu$ helps to make a clear distinction between AI and AD processes.
We also assume that the Gaussian process (possibly divided by a random variable to get a Student's \textit{t}) has a space-time correlation function, namely $\rho_{W^*}(h,k;\psi)= [1+(\|h\|/\psi_1)^2]^{-1} \, \exp(-k/\psi_2)$, for spatial lag $h$ and temporal lag $k$, which is a separable version of 
the space-time correlation function in \citet{Gneiting:2002}.
Some results with a different correlation function are presented in the Supplementary material.
The separability in the correlation function implies a parsimonious parameterization and speeds up the simulation of the processes.
Note that, even if $W^*(s,t)$ is a Gaussian process with separable covariance function, this is not the case for the non linearly transformed process $W(s,t)$, when its covariance function exists.

Model 1 and Model 3 share the same set of copula parameters, $\theta_D:=(\delta,\phi,\psi_1,\psi_2)$, but the role and interpretation of the values is specific to each model, especially for the $\delta$ parameter.

The training of the neural network involved, for both Model 1 and Model 3, $30\,000$ simulated datasets with the same $30$ spatial location and time dimension  of the observed one: $20$ years $\times\; 92$ days. For each dataset, a parameter value is uniformly generated on the following sets: $(0,1)$ for $\delta$, $(0,2.5)$ for $\phi$, $(4,16)$ for $\psi_1$, $(0,2.5)$ for $\psi_2$. 
For the parameters $\phi$, $\psi_1$ and $\psi_2$, the boundaries help to reduce the number of simulations needed to cover the four-dimensional parameter space. To set the boundaries, we checked that the resulting parameter space encompasses a range of values of $\chi(0.9)$, evaluated by Monte Carlo method at the observed distances in space and time, consistent with the range of the empirical values reported in Figure \ref{fig:chi_u_NL_space_time}.

To model the marginal distribution of the data, we follow the POT approach described in \ref{section:model_marginal}.
In particular, we assume that the thresholds $\mu(s,t)=\mu(s)$ in \eqref{eq:F_inv_GPD} depend only on the spatial location $s$.
This assumption is based on the fact that we restrict   the analysis to the spring months, for which we assume temporal stationarity of the data.
The value of $\mu(s)$ is preliminarily estimated by quantile regression for the $0.90$-quantile with spatial coordinates as covariates.
In addition, following \cite{bortot2024model}, we assume that the distribution of exceedances is invariant with spatial location, resulting in a scale parameter $\sigma$ and a shape parameter $\xi$ that are constant over the whole province of North Brabant. In Section \ref{section:suppl_marginal} in the Supplementary material, a justification for this choice is provided by fitting site-specific GPD models to the data and showing that, for all stations, the confidence intervals include the values of $\sigma$ and $\xi$ estimated on the overall data. This could be intuitively explained by the small size of the studied spatial domain and by its geographical regularity.

\subsection{Simulated data}\label{section:simulation_study}

The goal of the following simulation study is not to be exhaustive but rather to understand how a neural network-based estimation method for the copula model parameters, described in Section \ref{section:estimation_method}, might be able to infer the different classes of extremal dependence and to produce at least experimentally consistent estimates on a real-world example. In this regard, the spatial configuration was determined by selecting the pattern of sites in North Brabant with an equal number of observations over time. Furthermore, the models for temporal and spatio-temporal dependence remain the same as in Section \ref{section:application_model}.

\begin{figure}[htbp]
\begin{subfigure}[h]{0.5\linewidth}
\includegraphics[width=\linewidth]{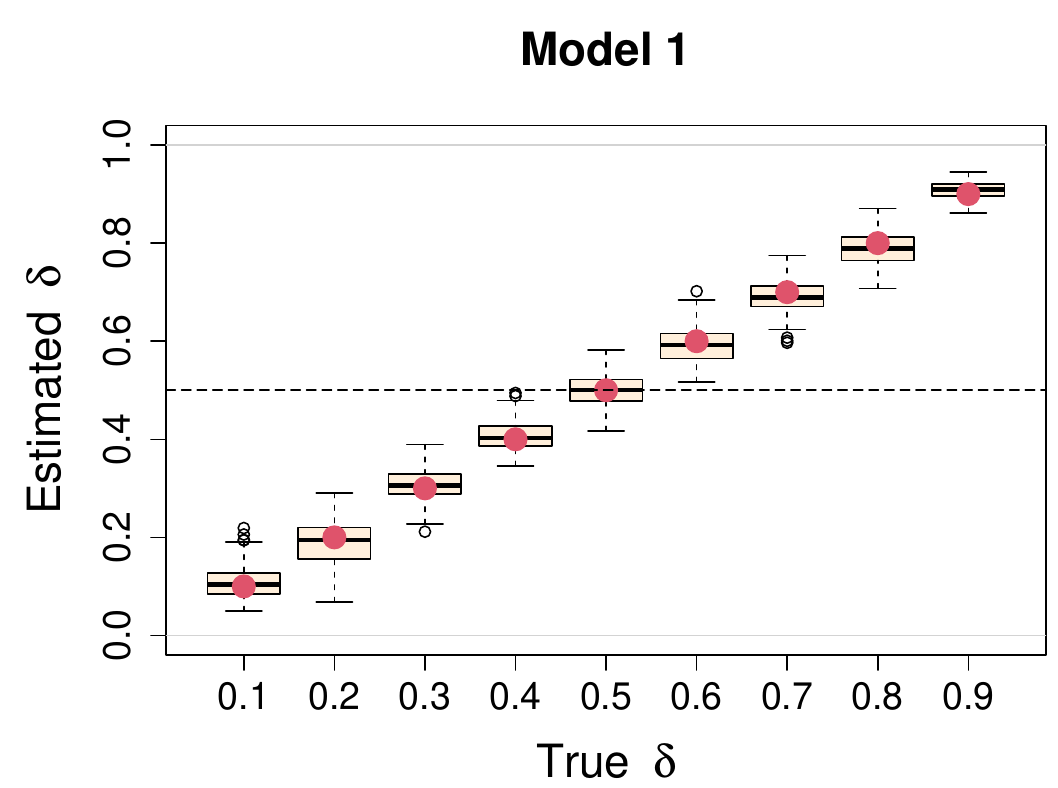}
\end{subfigure}
\hfill
\begin{subfigure}[h]{0.5\linewidth}
\includegraphics[width=\linewidth]{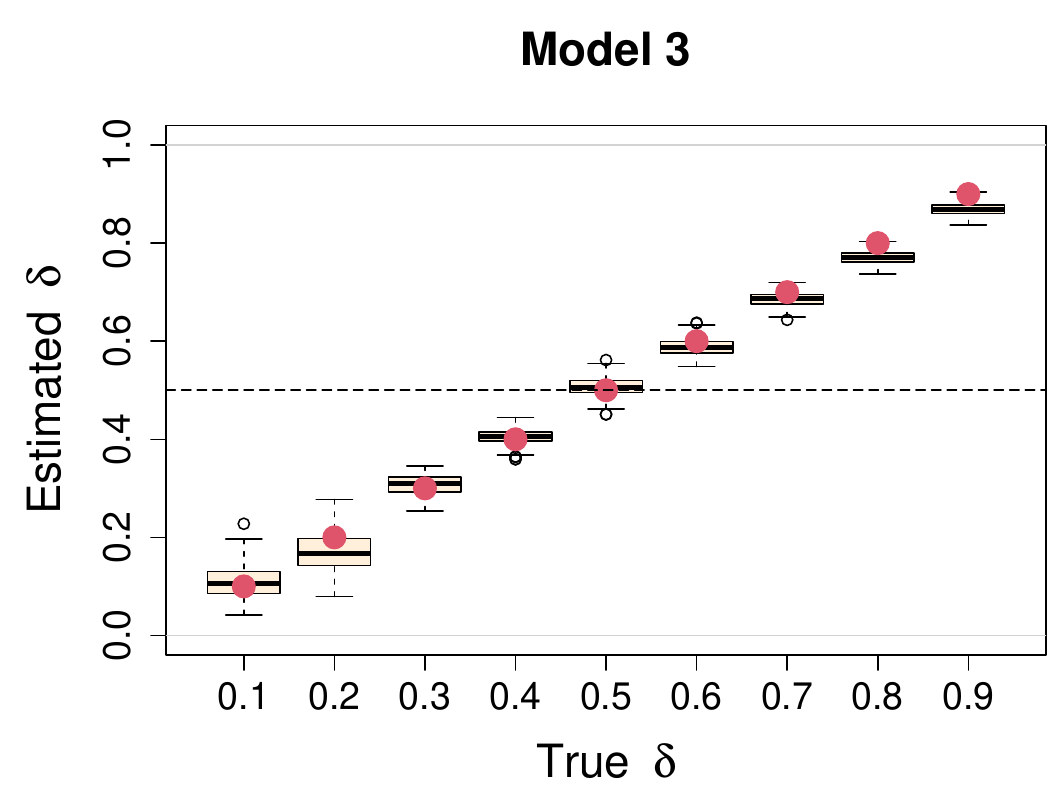}
\end{subfigure}%
\caption{True (red points) and estimated (boxplots) values of $\delta$ on 200 independently simulated datasets, for Model 1 (left) and Model 3 (right).}
\label{fig:sim_study_delta}
\end{figure}

Because of its relevance to the determination of the extremal dependence, we focus first on the estimation of $\delta$. 
For each value of $\delta$ in $\{0.1,0.2,\dots,0.9\}$, 200 data sets are generated from Models 1 and 3. In this experiment, we keep the true values of the other parameters constant and equal to the estimates displayed in Table \ref{table:estimated_values}, i.e.\ $\phi=0.874, \psi_1=9.107, \psi_2=0.328$ for Model 1, $\phi=1.045, \psi_1=10.045, \psi_2=0.377$ for Model 3.
The results, for $\delta$ only, are shown in Figure \ref{fig:sim_study_delta}. For simplicity, the results for the other parameters are not shown here, even though they are also estimated jointly with $\delta$.

The estimation method seems to work well enough to discriminate correctly between values smaller or greater than $0.5$, confirming that inference on $\delta$ is a viable means of determining the extremal dependence of the data, within the possibilities of each model (see Table \ref{table:possible_models}).
The boxplots for an analogue simulation study referring to Model 2 and Model 4 are reported in Figure \ref{fig:sim_study_delta_2_4} in the Supplementary material.

\begin{figure}[htbp]
\includegraphics[width=\linewidth]{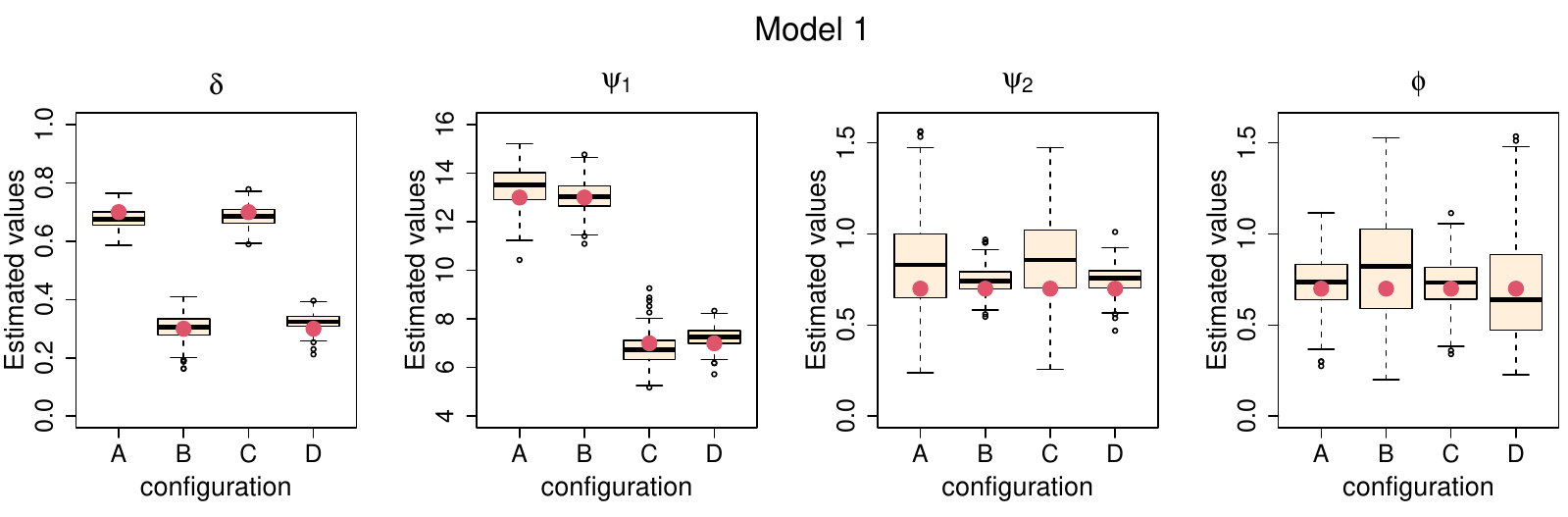}
\includegraphics[width=\linewidth]{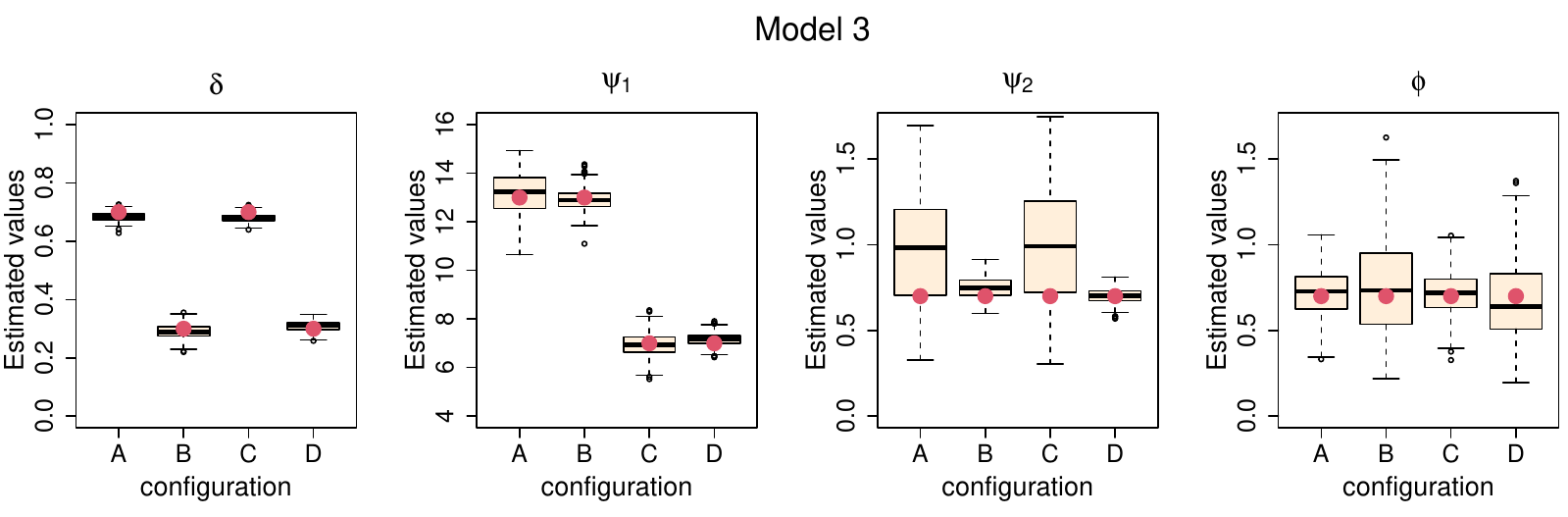}
\caption{Boxplots of the estimated values for $\delta=0.7$ (A,C) or $\delta=0.3$ (B,D), $\psi_1=13$ (A,B) or $\psi_1=7$ (C,D), $\psi_2=0.7$ and $\phi=0.7$ (red points) obtained on 400 simulated datasets from Model 1 and Model 3.}
\label{fig:sim_study_psi_phi}
\end{figure}

\begin{figure}[t]
\includegraphics[width=\linewidth]{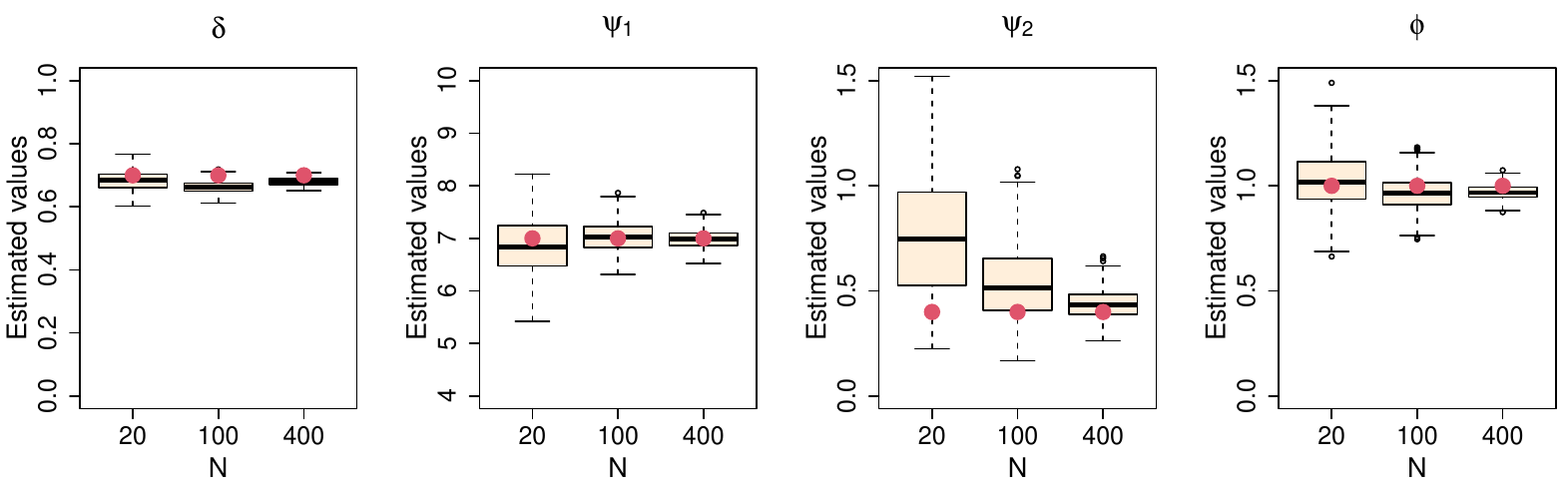}
\caption{Boxplots of estimated values for $\delta=0.7$, $\psi_1=7$, $\psi_2=0.4$ and $\phi=1$ (red points) obtained on 400 datasets, with $N\in\{20,100,400\}$ independent realizations from the space-time Model 1 in each dataset.}
\label{fig:sim_study_consistency}
\end{figure}

A second experiment, shown in Figure \ref{fig:sim_study_psi_phi} and Figure \ref{fig:sim_study_consistency}, focuses on the estimation of the other dependence parameters $\psi_1$, $\psi_2$ and $\phi$. Figure \ref{fig:sim_study_psi_phi} displays the boxplots of the estimated values for four configurations (A, B, C, D) of $\delta$, $\psi_1$ (scale of the spatial correlation of the Gaussian process  $W^*(s,t)$), $\psi_2$ (scale of the temporal correlation for the same process) and $\phi$ (scale of the temporal correlation of the Gaussian process  $R^*(t)$). The configurations are
\begin{itemize}
    \item A: $R(t)$ leads the mixture ($\delta=0.7$), with $\psi_1=13$, $\psi_2=0.7$ and $\phi=0.7$;
    \item B: $W(s,t)$ leads the mixture ($\delta=0.3$), with $\psi_1=13$, $\psi_2=0.7$ and $\phi=0.7$;
    \item C: $R(t)$ leads the mixture ($\delta=0.7$), with $\psi_1=7$, $\psi_2=0.7$ and $\phi=0.7$;
    \item D: $W(s,t)$ leads the mixture ($\delta=0.3$), with $\psi_1=7$, $\psi_2=0.7$ and $\phi=0.7$.
\end{itemize}
In all cases, spatial locations, time periods and number of independent replication (i.e., number of years) are fixed as the ones in the dataset. For simplicity, the values of $\psi_2$ and $\phi$ are the same in all configurations; note that other values are explored in Figure \ref{fig:sim_study_consistency}.
While the spatial dependence parameter $\psi_1$ is well estimated for both Model 1 and Model 3 (the corresponding boxplots referring to Model 2 and Model 4 are reported in Figure \ref{fig:sim_study_psi_phi_2_4} in the Supplementary material), the estimation of the temporal dependence parameters is more difficult in configurations A and C (for $\psi_2$) and B and D (for $\phi$), i.e.\ when the corresponding process is dominated by the other in the mixture.

Finally, we explore the consistency of the estimation method for $\psi$ and $\phi$ in Figure \ref{fig:sim_study_consistency}; here $\delta=0.7$, and the identifiability of $\psi_2$ is even more difficult because its value is lower than that of $\phi$, the time dependence parameter of the dominating process.
Indeed, when we fix $N$, the number of independent replications of the space-time dependent data, at $20$, as in Figure $\ref{fig:sim_study_psi_phi}$, the estimation of $\psi_2$ is poor. However, as we increase $N$ to $100$ and $400$, all the parameters are correctly identified by the estimation method presented in Section \ref{section:estimation_method}.

\subsection{Results for real data application}

The estimated values of the parameters and the bootstrap confidence intervals are reported in Table \ref{table:estimated_values}. In particular, $\hat{\delta}>0.5$ for both Model 1 and Model 3. Therefore, both models support the conclusion that rainfall data in the North Brabant province are asymptotically dependent in space and asymptotically independent in time (see Table \ref{table:possible_models}), which is consistent with the findings of \cite{bortot2024model} and with the preliminary analysis shown in Figure \ref{fig:chi_u_NL_space_time}.

\begin{table}[h!]
\begin{center}
\begin{tabular}{| c | c | c | c | c | c | c |} 
 \hline
 \multirow{2}{*}{Model} & \multicolumn{4}{c|}{Dependence parameters ($\theta_D$)} & \multicolumn{2}{c|}{Marginal parameters ($\theta_M$)} \\
  \cline{2-7} 
 & $\delta$ & $\phi$ & $\psi_1$ & $\psi_2$ & $\sigma$ & $\xi$ \\ 
 \hline
 \multirow{2}{*}{1} & 0.577 & 0.874 & 9.107 & 0.328 & 46.34 & 0.114 \\
 & \footnotesize(0.519,0.634) & \footnotesize(0.691,1.127) & \footnotesize(8.367,10.112) & \footnotesize(0.176,0.588) & \footnotesize(37.97,62.19) & \footnotesize(-0.026,0.211) \\
 \hline
 \multirow{2}{*}{3} & 0.631 & 1.045 & 10.045 & 0.377 & 46.34& 0.114 \\
 & \footnotesize(0.585,0.645) & \footnotesize(0.827,1.297) & \footnotesize(9.703,11.212) & \footnotesize(0.204,0.632) & \footnotesize(38.13,62.07) & \footnotesize(-0.026,0.204) \\
 \hline
\end{tabular}
\caption{Estimated values and $0.90$ bootstrapped confidence intervals, in parentheses, for the parameters in $\theta_D$ and $\theta_M$.}
\label{table:estimated_values}
\end{center}
\end{table}

\begin{figure}[ht!]
\begin{subfigure}{0.329\linewidth}
\includegraphics[width=\linewidth]{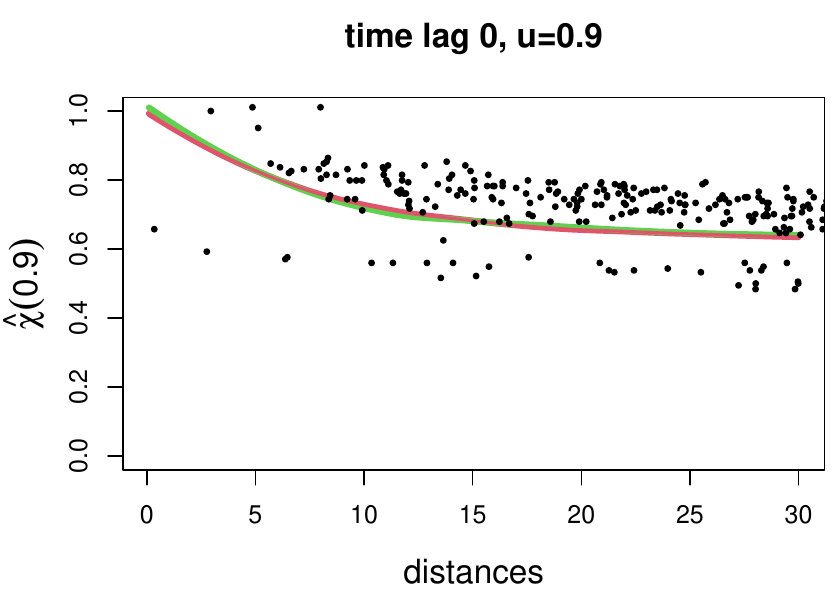}
\end{subfigure}
\begin{subfigure}{0.329\linewidth}
\includegraphics[width=\linewidth]{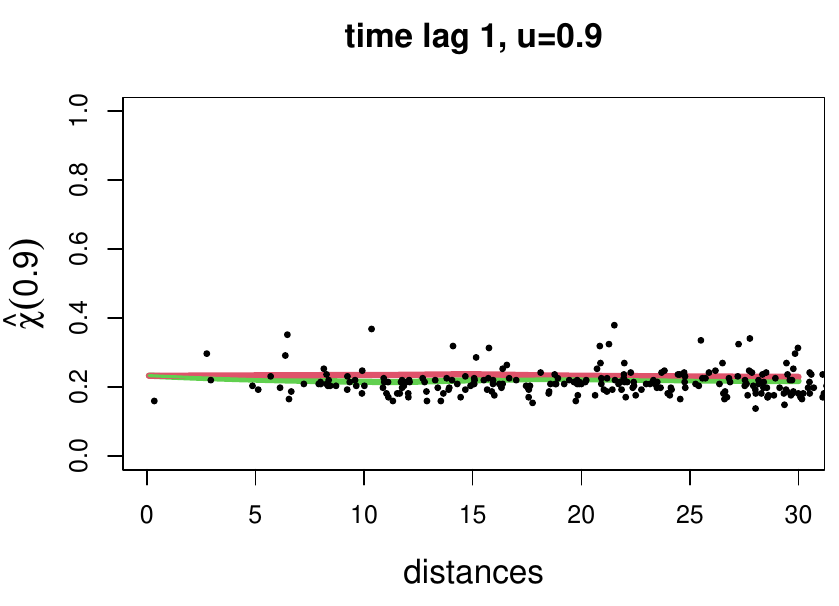}
\end{subfigure}
\begin{subfigure}{0.329\linewidth}
\includegraphics[width=\linewidth]{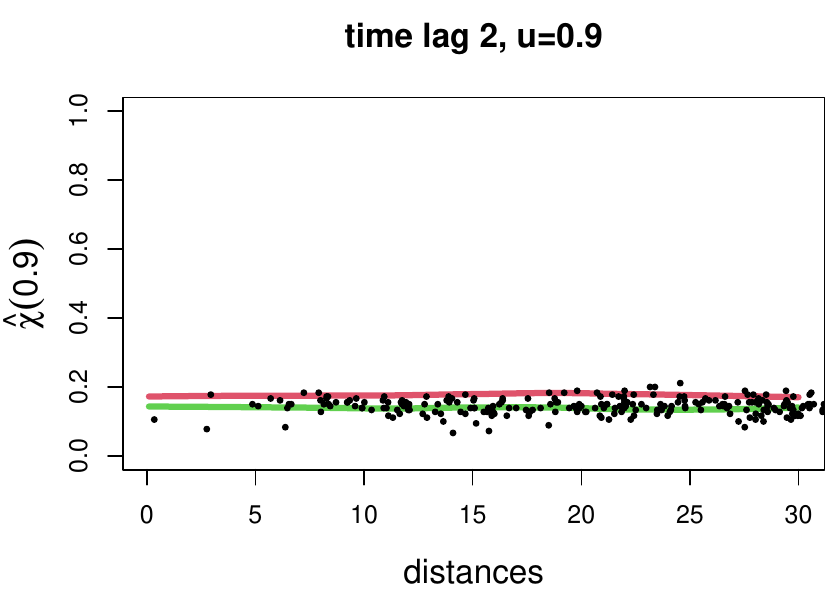}
\end{subfigure}

\begin{subfigure}{0.329\linewidth}
\includegraphics[width=\linewidth]{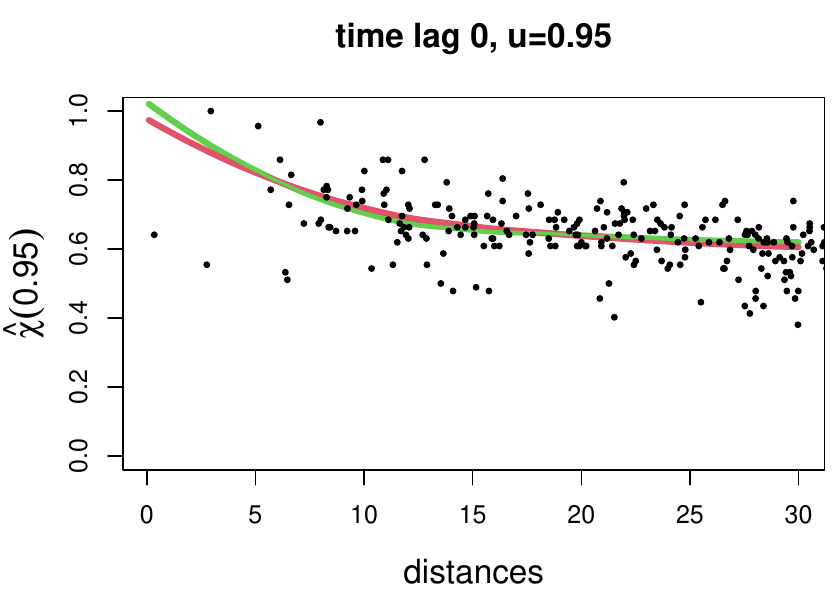}
\end{subfigure}
\begin{subfigure}{0.329\linewidth}
\includegraphics[width=\linewidth]{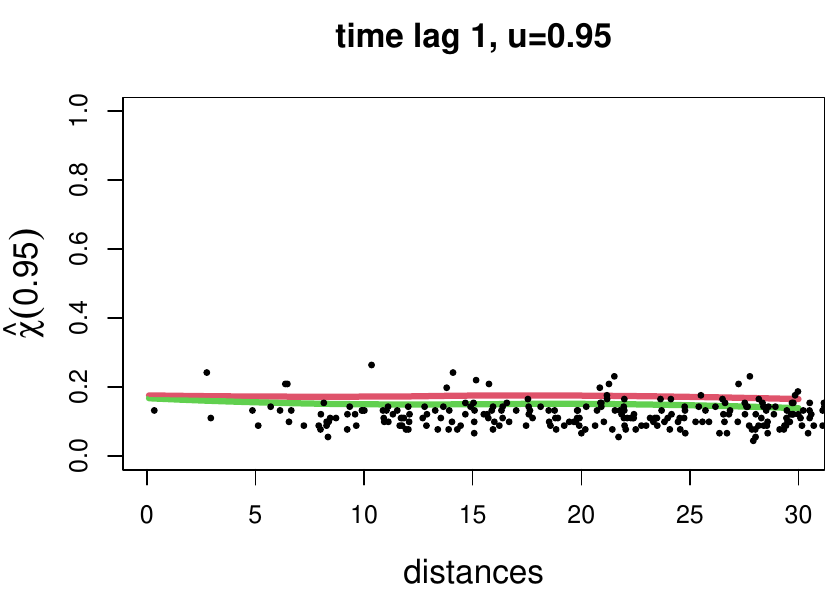}
\end{subfigure}
\begin{subfigure}{0.329\linewidth}
\includegraphics[width=\linewidth]{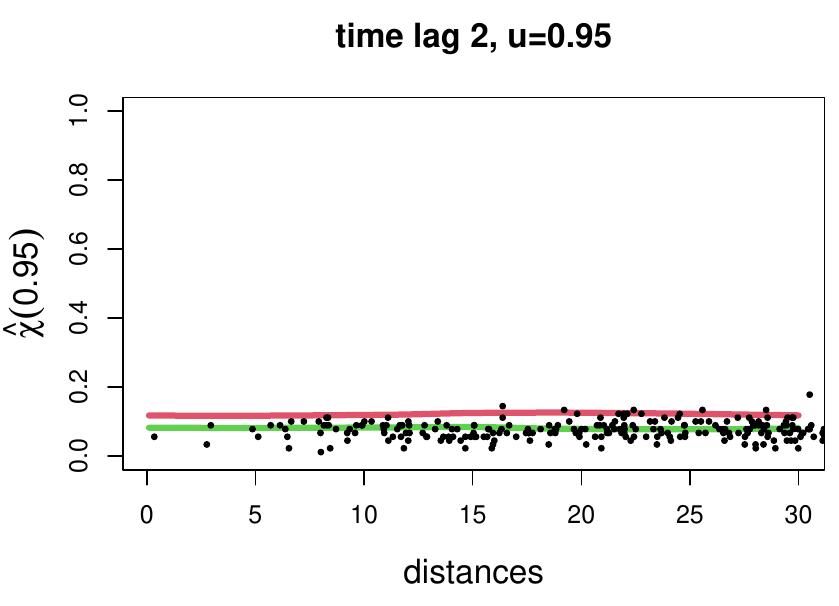}
\end{subfigure}

\begin{subfigure}{0.329\linewidth}
\includegraphics[width=\linewidth]{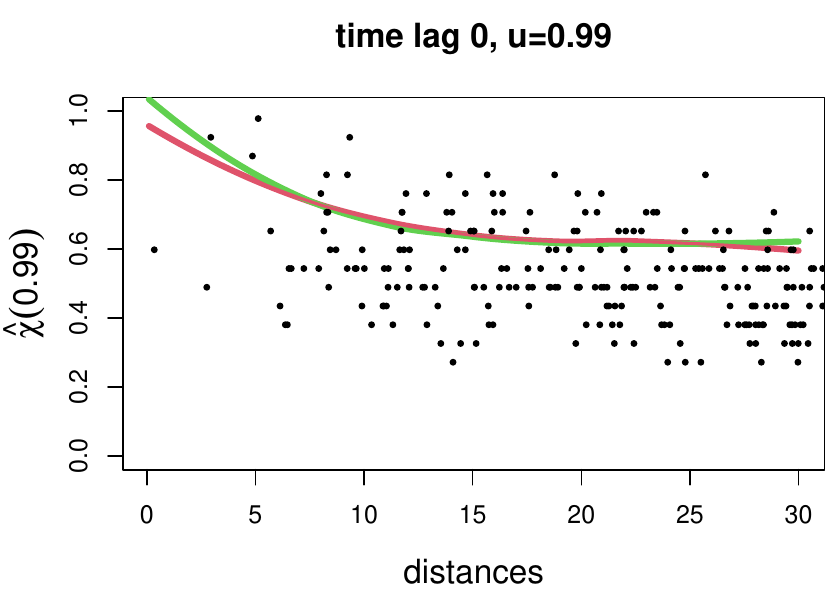}
\end{subfigure}
\begin{subfigure}{0.329\linewidth}
\includegraphics[width=\linewidth]{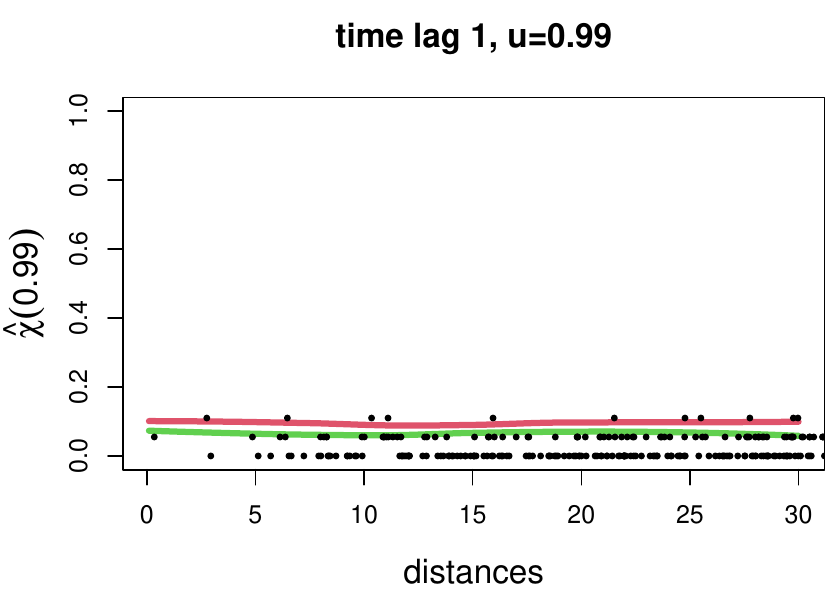}
\end{subfigure}
\begin{subfigure}{0.329\linewidth}
\includegraphics[width=\linewidth]{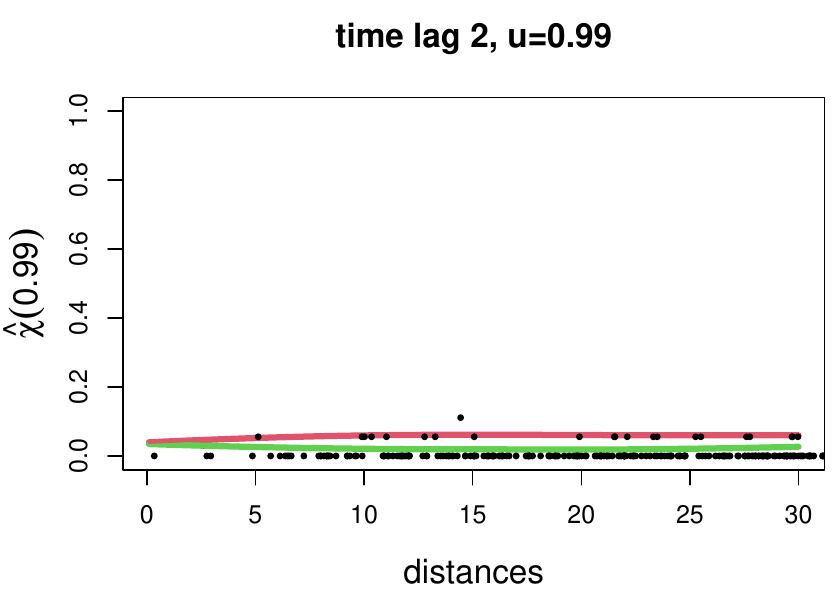}
\end{subfigure}
\caption{Values of  ${\chi(u)}$  for different spatial distances and temporal lags, derived from the estimated Model 1 (red lines), from the Model 3 (green lines) and the empirical ones for each pair of stations (dots).}
\label{fig:diagn_mod_1_3}
\end{figure}

Figure \ref{fig:diagn_mod_1_3} shows the empirical pairwise coefficient $\hat{\chi}_{1,2}(u)$, as defined in \eqref{empirical_chi}, for pairs $(Y_1,Y_2)$ at different distances in space and time and for $u\in\{0.90,0.95,0.99\}$. The dots represent the empirical coefficients observed in the dataset, while the lines are smoothed estimates from the two models, evaluated by simulation. Both the models are able to capture the decay of joint tail probabilities, characterized by the persistence of positive values in space, symptom of asymptotic dependence (first column), and by values rapidly going to zero for pairs at time lag 1 or 2, that are asymptotically independent (second and third columns). Figure \ref{fig:diagn_mod_1_3_ExpGaussian} in the Supplementary material shows the same plot for a model defined with correlation functions different from those presented in \ref{section:application_model}.

It is not straightforward to determine which model best fits the data from Figure \ref{fig:diagn_mod_1_3}.
To formally do so, we perform a likelihood-free model selection based on cross validation.
Since the dataset includes twenty years of daily observation, and we assume independence and stationarity among years, we split it into a $15$-years block and a $5$-years block, randomly sampled. The grids of empirical $\hat{\chi}$ described in Section \ref{section:estimation_method} are computed separately on the two blocks. The ones computed on the $15$-years block are used to get parameter estimates for each model from the neural network, and $500$ 5-years datasets are simulated to get a Monte Carlo estimate of the grids, to be compared with those computed on the $5$-years left-out block, through the Root Mean Squared Error (RMSE). This procedure is repeated iteratively $50$ times. The resulting RMSE is equal to $0.066$ for Model 1 and to $0.064$ for Model 3, indicating a slight preference for the latter.

Another diagnostic measure that we adopt, similarly to \cite{Bacro:Gaetan:Opitz:Toulemonde:2020}, is
$$\chi^*_{s_i;k}(u):=\Pr\left(Y(s_j,t)>\hat{F}_j^{-1}(u),\forall s_j\in\partial s_i\mid Y(s_i,t-k)>\hat{F}_i^{-1}(u)\right)$$
where $\partial s_i$ is the set of the four nearest neighbors of site $s_i$, $i=1,\dots,30$ and $\hat{F}_i^{-1}(u)$ and $\hat{F}_j^{-1}(u)$ are the estimated quantiles of order $u$ for the locations $s_i$ and $s_j$.
This allows to evaluate the performance of the models with respect to higher-dimensional extremal dependence in space and time, while the parameter estimation relies on pairwise information only, see \eqref{empirical_chi}.
The empirical estimates of $\chi^*_{s_i;k}(u)$, $\hat{p}_i(k,u)$, for $k\in\{0,1,2\}$ and $u\in\{0.90,0.95\}$, are compared with parametric bootstrap estimates $\tilde{p}^{(j)}_i(k,u)$, for $j=1,\dots,1000$, through the site-specific RMSE
$$\text{RMSE}_i(k,u)=\left\{\frac{\sum_{j=1}^{1000} (\tilde{p}^{(j)}_i(k,u)-\hat{p}_i(k,u))^2}{1000}\right\}^{1/2}$$
and the total mean $\text{RMSE}(k,u)=\sum_{i=1}^{30}\text{RMSE}_i(k,u)/30$, which is reported in Table \ref{table:RMSE}. Model 3 seems to perform better than Model 1 on the multivariate spatial dependence for neighbors at the same time periods ($k=0$) and higher quantile ($u=0.95$), while the performance of the two models is similar in the other settings.
Table \ref{table:RMSE_corGau} in the Supplementary material contains the same measures for a model defined with correlation functions different from those presented in \ref{section:application_model}.

\begin{figure}[t]
    \centering
    \includegraphics[width=1\linewidth]{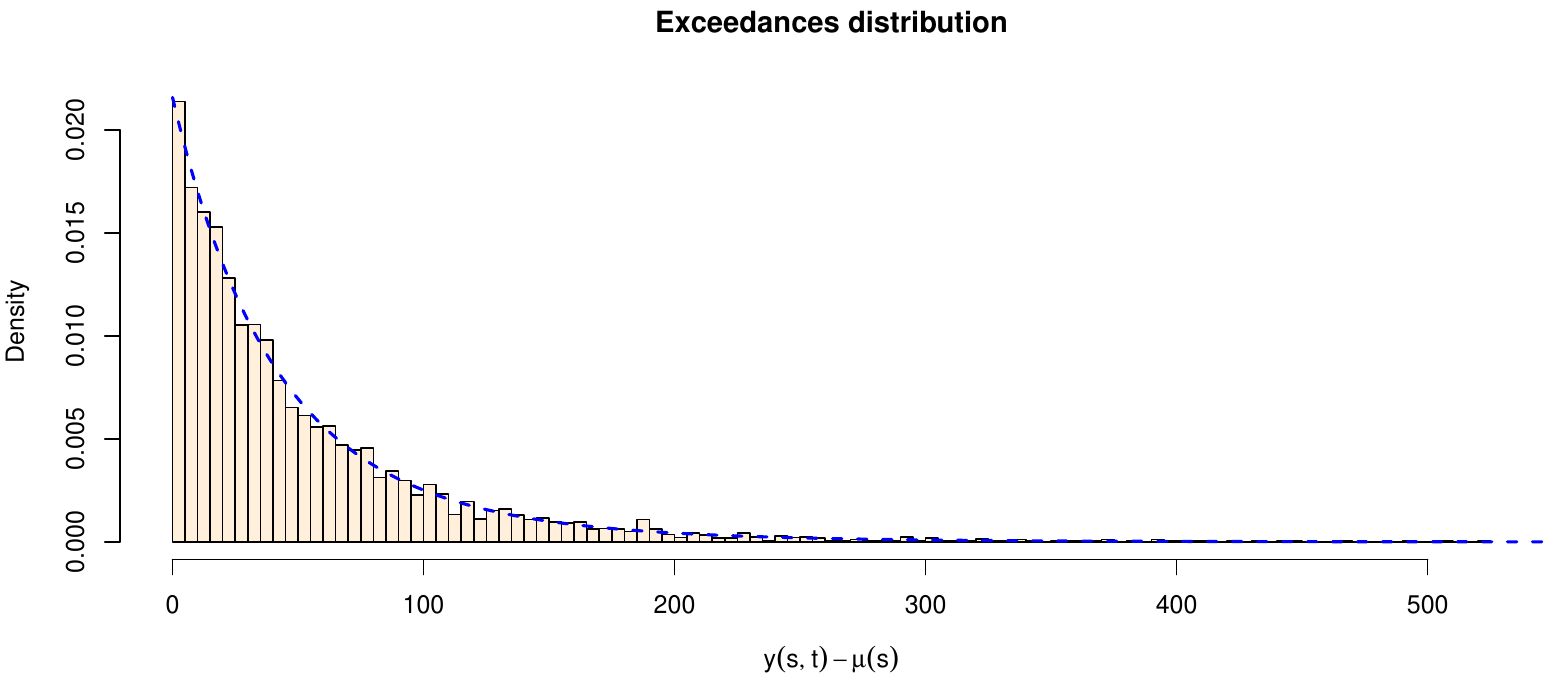}
\caption{Observed right tail (threshold exceedances) of the marginal distribution. The dashed blue line is the GPD density with parameters equal to the maximum likelihood estimates $\hat{\sigma}=46.34$ and $\hat{\xi}=0.114$.}
\label{figure:diagn_mar}
\end{figure}

\begin{table}[h!]
\begin{center}
\begin{tabular}{| c | c | c | c | c | c | c |} 
 \hline
  \multirow{2}{*}{Model} & \multicolumn{2}{c|}{$k=0$} & \multicolumn{2}{c|}{$k=1$} & \multicolumn{2}{c|}{$k=2$} \\
  \cline{2-7} 
   & $u=0.90$ & $u=0.95$ & $u=0.90$ & $u=0.95$ & $u=0.90$ & $u=0.95$ \\ [0.5ex] 
 \hline
 1 & 0.116 & 0.138 & 0.025 & 0.027 & 0.027 & 0.028 \\
 \hline
 3 & 0.111 & 0.109 & 0.028 & 0.030 & 0.027 & 0.030 \\
 \hline
\end{tabular}
\caption{$\text{RMSE}(k,u)$ between empirical estimates of $\chi^*_{s_i;k}(u)$ and estimates from Model 1 and Model 3 (see Table \ref{table:possible_models}).}
\label{table:RMSE}
\end{center}
\end{table}

Figure \ref{figure:diagn_mar} shows the distribution of thresholds exceedances observed in the rainfall dataset. The thresholds $\mu(s)$ are the $0.90$-quantiles   at each station, estimated via quantile regression.
The dashed blue line is the theoretical GPD density with the parameter values set as the maximum likelihood estimate.
The estimated values and the bootstrap confidence intervals for the scale parameter $\sigma$ and for the shape parameter $\xi$ are reported in Table \ref{table:estimated_values}.

\begin{figure}[t!]
\begin{subfigure}{0.505\linewidth}
\includegraphics[width=\linewidth]{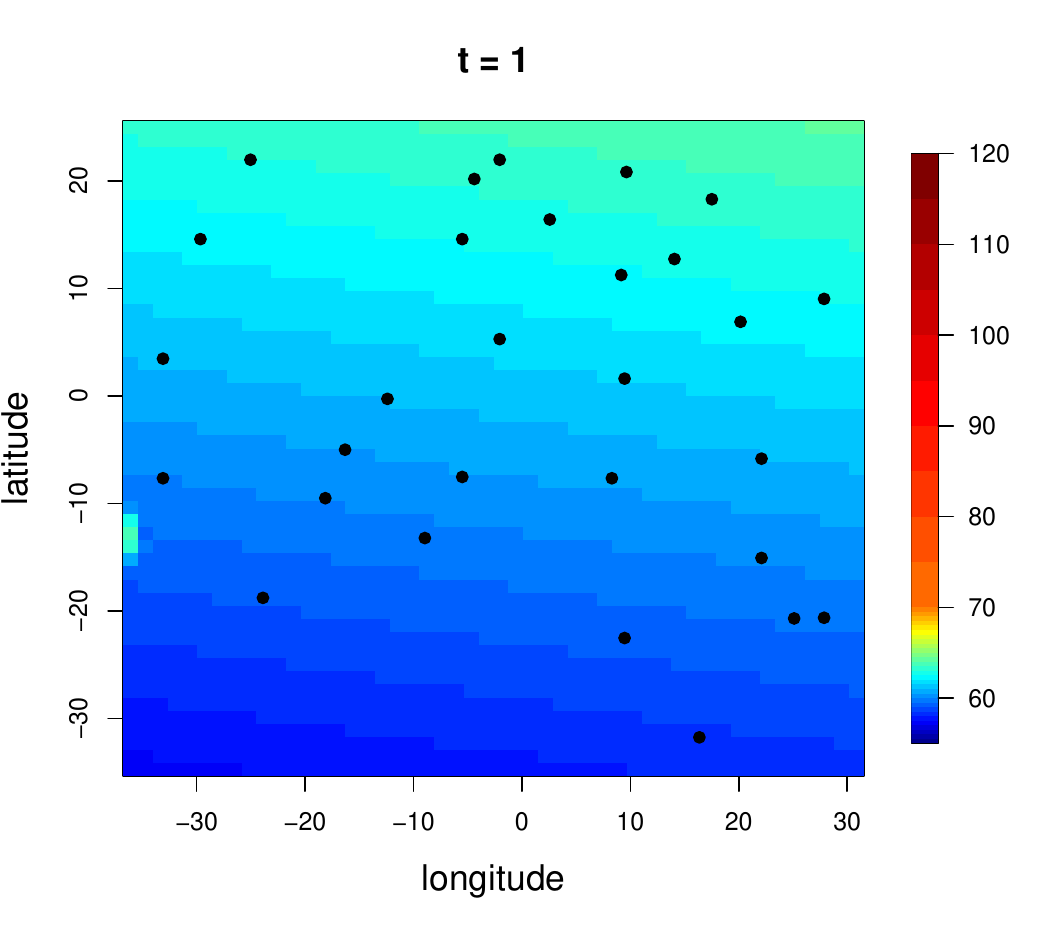}
\end{subfigure}
\begin{subfigure}{0.505\linewidth}
\includegraphics[width=\linewidth]{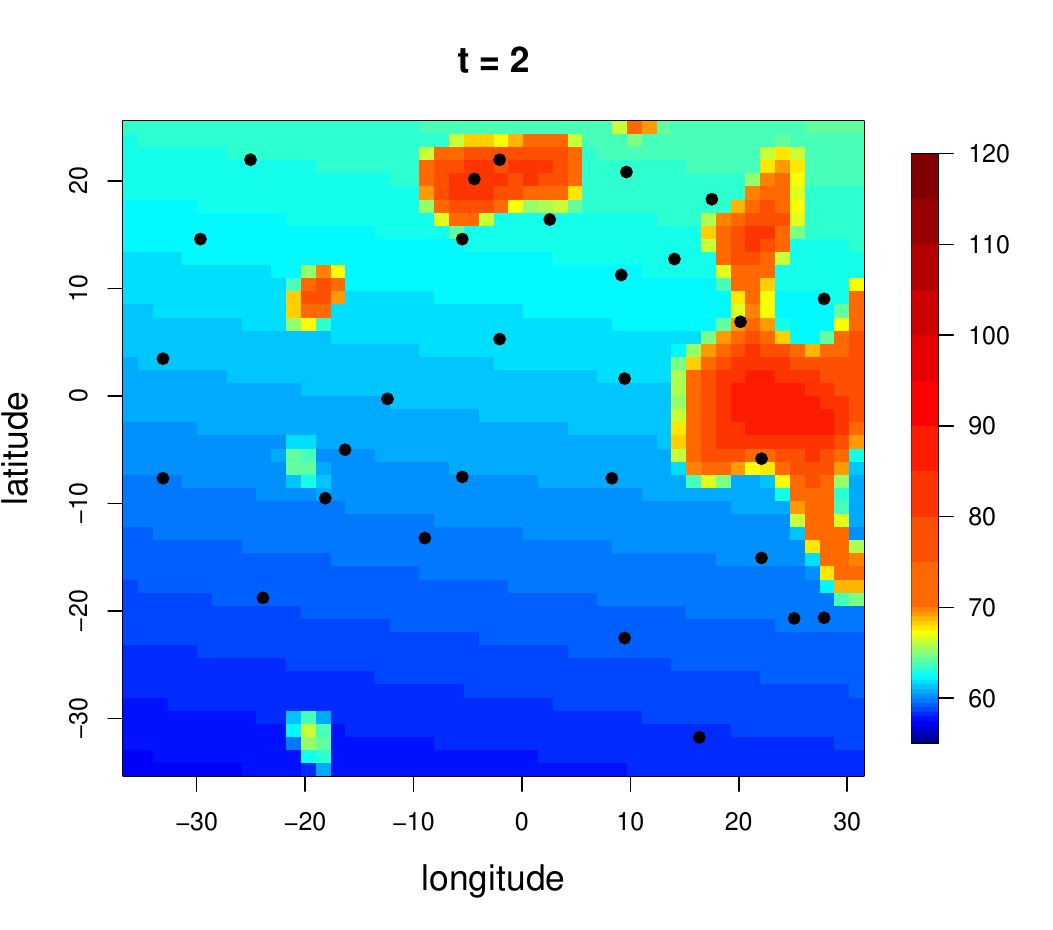}
\end{subfigure}

\begin{subfigure}{0.505\linewidth}
\includegraphics[width=\linewidth]{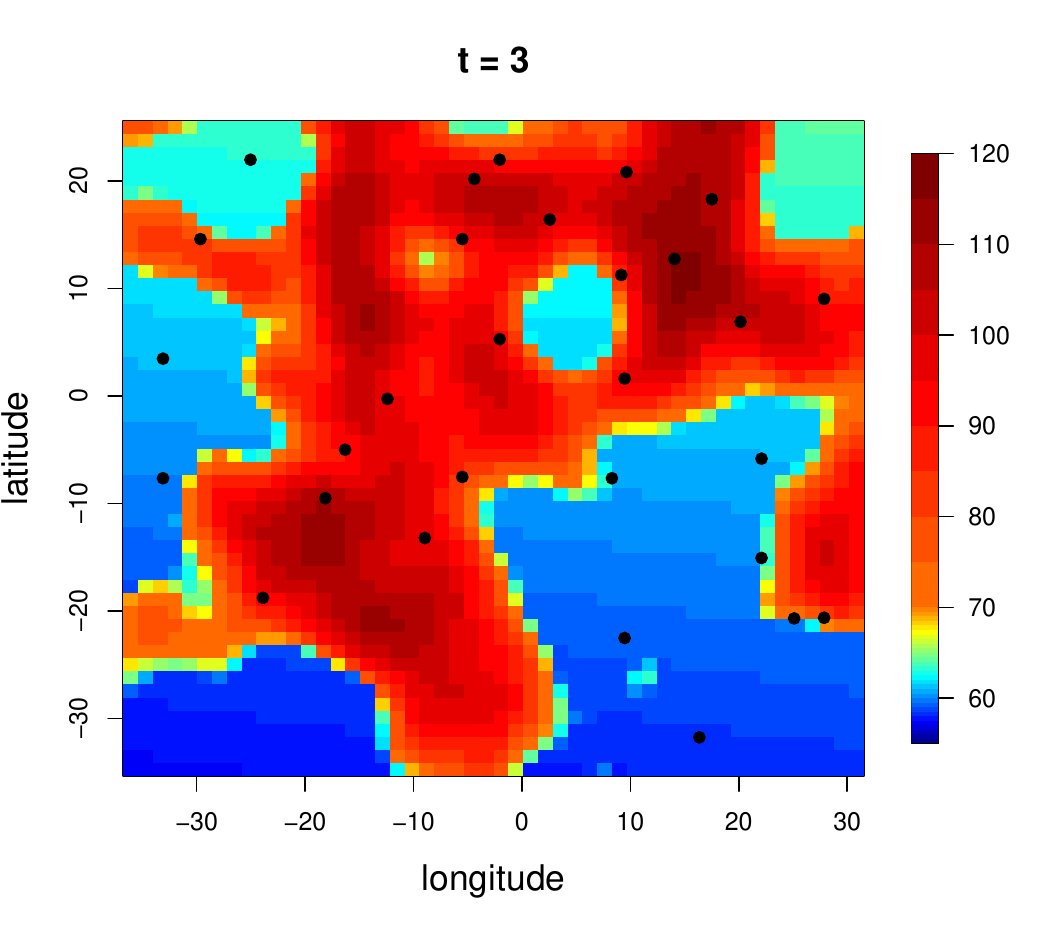}
\end{subfigure}
\begin{subfigure}{0.505\linewidth}
\includegraphics[width=\linewidth]{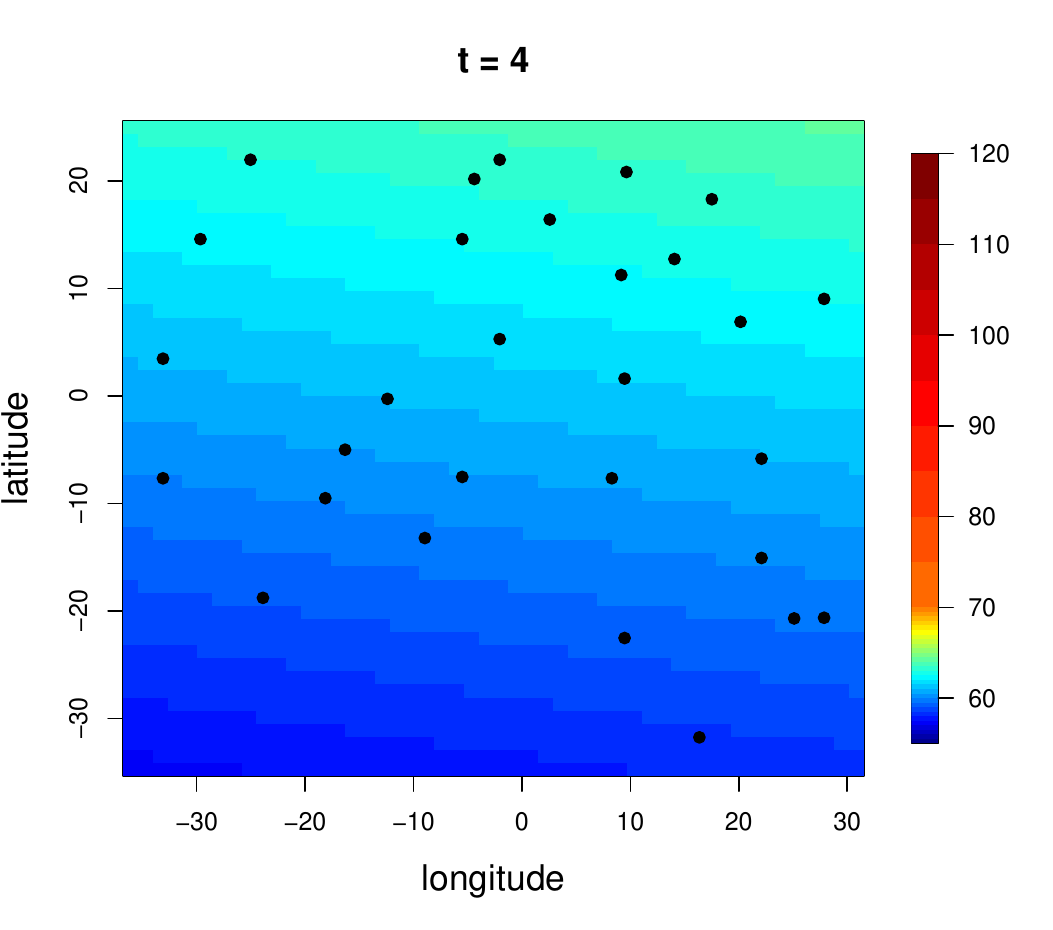}
\end{subfigure}
\caption{A simulation of four consecutive days of rainfall from Model 1 (see Table \ref{table:possible_models}) with $\delta>0.5$: AD in space and AI in time. Data exceeding the 0.90 marginal quantile are highlighted in red, while the rest of the data is censored and the estimated 0.90 quantiles are displayed. Black points denote the stations observed in the data application.}
\label{figure:simulated_storm}
\end{figure}

Finally, Figure \ref{figure:simulated_storm} shows a simulation of four consecutive days of rainfall from Model 1 with the estimated values of the parameters: $\delta=0.577$, $\phi=0.874$, $\psi_1=9.107$, $\psi_2=0.328$, $\sigma=46.34$, $\xi=0.114$. The simulated data below the 0.90 marginal quantile are censored and the value of the thresholds, which are spatially varying, are colored in blue and green. The areas colored in red and yellow are those with rainfall amounts exceeding the thresholds. Black points denote the stations observed in the data application. Since $\delta>0.5$, the model is AD in space and AI in time: this results in a storm with extreme amounts of rainfall over the majority of the spatial domain at $t=3$, but rapidly fading away in the following time period.

\section{Conclusions}\label{section:conclusions}
It is widely acknowledged in the literature that sub-asymptotic models should be employed to capture the dependence between values exceeding a convenient high threshold \citep{huser2024modelingspatialextremesenvironmental}. However, the use of this type of models has mainly focused on spatial dependence, while spatio-temporal dependence is less studied. Indeed, the presence of a distinct type of extremal dependence in the spatial and temporal dimensions renders specification challenging.
In comparison to previous attempts that have sought to achieve the same objective \citep{bortot2024model}, this work proposes a solution that has the advantage of being able to capture the different types of dependence through a parameter whose value can be estimated from the data. The estimation of this parameter, as well as the other model parameters, is achieved through the utilization of the computational power offered by modern numerical libraries for learning neural networks. 

The proposed spatio-temporal model exhibits an extremal spatio-temporal dependence that is stationary within the specified period. In consideration of its potential application to the study of climate change, a logical subsequent step is to consider a non-stationary model, similarly to the approach outlined in \cite{Maume:Ribereau:Zeidan:2024}.
Moreover, in order to apply the model to larger spatial domains, the introduction of spatial non-stationarity would be an useful extension.

The flexibility of the models presented in this work makes them suitable for analyzing many type of data whose extremal dependence in space and time is of interest, and notably environmental data. Nevertheless, given the novelty of the employed parameter estimation method, a greater focus is needed on competitive model selection, which would require the development of new diagnostic tools, and will be the subject of future research.

\section*{Acknowledgments}
\noindent
The second author received support from the Geolearning Research Chair, which is a joint initiative between Mines Paris and the French National Institute for Agricultural Research (INRAE).
The authors thank Thomas Opitz for helpful discussions and suggestions about the random scale construction and Rapha{\"e}l Huser for hints about the estimation method.
 The authors also thank the anonymous reviewers for their constructive suggestions that helped improve the quality of the manuscript.
\bibliography{paper}
\bibliographystyle{apalike-citesort}

\appendix
\section{Proof on dependence properties (pairs in time and space-time)}\label{section:appendix_time_spacetime}

This appendix contains the proofs on the dependence properties of model \eqref{model_X_RW} stated in Section \ref{section:dependence_properties}, for pairs in time and space-time. Pairs in space are covered by \ref{section:appendix_space}.
Let $(X_1,X_2)\,=(R_1^{\delta}W_1^{1-\delta},R_2^{\delta}W_2^{1-\delta})$, with $X_{i}:=X(s_i,t_i)$, $R_i:=R(t_i)$ and $W_{i}:=W(s_i,t_i)$, for $i=1,2$, such that $t_1\ne t_2$, be a generic pair from model \eqref{model_X_RW}.
To study the extremal dependence of $(X_1,X_2)$, we have to compute their coefficient $\chi_X$, equivalent to the one defined in \eqref{def:chi_X_12}, and in particular its numerator, for a fixed $u$, $\Pr\left(X_1>F^{-1}_{X_1}(u),X_2>F^{-1}_{X_2}(u) \right)$.

\bigskip
\subsection*{Case 1: \texorpdfstring{$\delta<0.5$}{delta<1/2} (third column of Table \ref{table:possible_models})}
\medskip

\noindent$(X_1,X_2)$ belong to the same extremal dependence class as $(W_1,W_2)$, i.e.,

\begin{itemize}
\item[-]  $(X_1,X_2)\; \text{are AI} \qquad\qquad\qquad\quad\;\;\; \text{if}\; (W_1,W_2) \;\text{are AI, with}\; \eta_W<1$;
\item[-]  $(X_1,X_2)\; \text{are AD} \qquad\qquad\qquad\quad\:\: \text{if}\; (W_1,W_2) \;\text{are AD}$.
\end{itemize}
\bigskip
\underline{Proof}:
By Breiman's lemma (\citealp{breiman1965some}, see also \citealp{engelke2019extremal}, Lemma 8), as $x\rightarrow\infty$,
$$\Pr(X_1>x)=\; \Bar{F}_{X_1}(x)\sim \E\left(R_1^{\delta/(1-\delta)}\right) \Bar{F}_{W_1^{1-\delta}}(x)=\frac{1-\delta}{1-2\delta}\;x^{-1/(1-\delta)},$$
and, as $u\rightarrow1$ (see \citealp{engelke2019extremal}, eq.\ 22),
$$F^{-1}_{X_1}(u) \sim \E\left(R_1^{\delta/(1-\delta)}\right)^{1-\delta}F^{-1}_{W_1^{1-\delta}}(u)=\left(\frac{1-\delta}{1-2\delta}\right)^{1-\delta} \left(\frac{1}{1-u}\right)^{1-\delta}.$$
Therefore, as $u\rightarrow1$, the numerator of $\chi_X$ is
\begin{equation*}
\begin{split}
    \Pr\left(X_1>F^{-1}_{X_1}(u),X_2>F^{-1}_{X_2}(u) \right) \sim &\; \Pr\left(X_1>\left(\frac{1-\delta}{1-2\delta}\right)^{1-\delta} \left(\frac{1}{1-u}\right)^{1-\delta},X_2>\left(\frac{1-\delta}{1-2\delta}\right)^{1-\delta} \left(\frac{1}{1-u}\right)^{1-\delta} \right)\\
    =&\; \Pr\left(R_1^{\delta/(1-\delta)} W_1>\frac{1-\delta}{(1-2\delta)(1-u)},R_2^{\delta/(1-\delta)} W_2>\frac{1-\delta}{(1-2\delta)(1-u)} \right)\\
    <&\; \Pr\left(R_1^{\delta/(1-\delta)} W_1>\frac{1}{1-u},R_2^{\delta/(1-\delta)} W_2>\frac{1}{1-u} \right)\\
    =&\; \Pr\left(\min\left\{R_1^{\delta/(1-\delta)} W_1,R_2^{\delta/(1-\delta)} W_2\right\}>\frac{1}{1-u} \right)\\
    <&\; \Pr\left(R_1^{\delta/(1-\delta)}\min\{W_1,W_2\}>\frac{1}{1-u} \right) +\\
    &+ \Pr\left(R_2^{\delta/(1-\delta)}\min\{W_1,W_2\}>\frac{1}{1-u} \right)\\
    \sim& \;2\cdot(1-u)^{\min\left\{(1-\delta)/\delta,\;1/\eta_W\right\}},
\end{split}
\end{equation*}
where the last line derives from Breiman's Lemma, noting that, as $u\rightarrow1$, $\Bar{F}_{\min\{W_1,W_2\}}(x)\sim x^{-1/\eta_W}$, while $\Bar{F}_{R_i^{\delta/(1-\delta)}}(x)\sim x^{-(1-\delta)/\delta}$, for $i=1,2$.
Then, if $(W_1,W_2)$ are AI with $\eta_W<1$, $\min\left\{(1-\delta)/\delta,\;1/\eta_W\right\}>1$ and
$$\chi_X=\lim_{u\rightarrow1}\frac{\Pr\left(X_1>F^{-1}_{X_1}(u),X_2>F^{-1}_{X_2}(u) \right)}{1-u} \le \lim_{u\rightarrow1}\frac{(1-u)^{\min\left\{\frac{1-\delta}{\delta},\frac{1}{\eta_W}\right\}}}{1-u}=0,$$
i.e. $(X_1,X_2)$ are AI, with $\eta_X\le \max\left\{\delta/(1-\delta),\;\eta_W\right\}<1$.\vspace{0.3cm}\\
On the other hand,
\begin{equation*}
\begin{split}
    \Pr\left(X_1>F^{-1}_{X_1}(u),X_2>F^{-1}_{X_2}(u) \right) \sim &\;  \Pr\left(X_1>\left(\frac{1-\delta}{1-2\delta}\right)^{1-\delta} \left(\frac{1}{1-u}\right)^{1-\delta},X_2>\left(\frac{1-\delta}{1-2\delta}\right)^{1-\delta} \left(\frac{1}{1-u}\right)^{1-\delta} \right)\\
    =&\; \Pr\left(R_1^{\delta/(1-\delta)} W_1>\frac{1-\delta}{(1-2\delta)(1-u)},R_2^{\delta/(1-\delta)} W_2>\frac{1-\delta}{(1-2\delta)(1-u)} \right)\\
    =&\; \Pr\left(\min\left\{R_1^{\delta/(1-\delta)} W_1,R_2^{\delta/(1-\delta)} W_2\right\}>\frac{1-\delta}{(1-2\delta)(1-u)} \right)\\
    \ge&\; \Pr\left(\min\{R_1,R_2\}^{\delta/(1-\delta)}\;\min\{W_1,W_2\}>\frac{1-\delta}{(1-2\delta)(1-u)} \right)\\
    \sim&\; c(\delta)\, (1-u)^{\min\left\{(1-\delta)/(\delta\cdot\eta_R),\; 1/\eta_W\right\}},
\end{split}
\end{equation*}
with $c(\delta)>0$  for all $\delta<0.5$. If $(W_1,W_2)$ are AD, $\eta_W=1$, so $\min\left\{(1-\delta)/(\delta\cdot\eta_R),\; 1/\eta_W\right\}=1$ and
$$\chi_X=\lim_{u\rightarrow1}\frac{\Pr\left(X_1>F^{-1}_{X_1}(u),X_2>F^{-1}_{X_2}(u) \right)}{1-u} \ge \lim_{u\rightarrow1}\frac{c(\delta)(1-u)}{1-u}>0,$$
i.e. $(X_1,X_2)$ are AD.

\bigskip
\subsection*{Case 2: \texorpdfstring{$\delta>0.5$}{delta>1/2} (first column of Table \ref{table:possible_models})}
\bigskip

\noindent $(X_1,X_2)$ belong to the same extremal dependence class as $(R_1,R_2)$, i.e.,

\begin{itemize}
    \item[-] $(X_1,X_2)\; \text{are AI} \qquad\qquad\qquad\quad\;\;\; \text{if}\; (R_1,R_2) \;\text{are AI, with}\; \eta_W<1$;
\item[-]  $(X_1,X_2)\; \text{are AD} \qquad\qquad\qquad\quad\:\: \text{if}\; (R_1,R_2) \;\text{are AD}$.
\end{itemize}

\bigskip
\underline{Proof}: 
By Breiman's lemma,
$$\Pr(X_1>x)=\; \Bar{F}_{X_1}(x)\sim \E\left(W_1^{(1-\delta)/\delta}\right) \Bar{F}_{R_1^\delta}(x)=\frac{\delta}{2\delta-1}x^{-1/\delta},$$ 
$$F^{-1}_{X_1}(u) \sim \E\left(W_1^{(1-\delta)/\delta}\right)^\delta F^{-1}_{R_1^\delta}(u)=\left(\frac{\delta}{2\delta-1}\right)^\delta\left(\frac{1}{1-u}\right)^\delta.$$
Therefore, similarly to Case 1, as $u\rightarrow1$ the numerator of $\chi_X$ is
\begin{equation*}
\begin{split}
    \Pr\left(X_1>F^{-1}_{X_1}(u),X_2>F^{-1}_{X_2}(u) \right) \sim &\; \Pr\left(X_1>\left(\frac{\delta}{2\delta-1}\right)^\delta\left(\frac{1}{1-u}\right)^\delta,X_2>\left(\frac{\delta}{2\delta-1}\right)^\delta\left(\frac{1}{1-u}\right)^\delta \right)\\
    <&\; \Pr\left(R_1 W_1^{(1-\delta)/\delta}>\frac{1}{1-u},R_2 W_2^{(1-\delta)/\delta}>\frac{1}{1-u} \right)\\
    <&\; \Pr\left(\min\{R_1,R_2\}\,W_1^{(1-\delta)/\delta}>\frac{1}{1-u} \right) +\\
    &+\Pr\left(\min\{R_1,R_2\}\,W_2^{(1-\delta)/\delta}>\frac{1}{1-u} \right)\\
    \sim& \;2\cdot(1-u)^{\min\left\{(\delta/(1-\delta),\;1/\eta_R\right\}},
\end{split}
\end{equation*}
where the last line derives from Breiman's Lemma, noting that, as $u\rightarrow1$, $\Bar{F}_{\min\{R_1,R_2\}}(x)\sim x^{-1/\eta_R}$, while $\Bar{F}_{W_i^{(1-\delta)/\delta}}(x)\sim x^{-\delta/(1-\delta)}$, for $i=1,2$.
Then, if $(R_1,R_2)$ are AI with $\eta_R<1$, $\min\left\{(\delta/(1-\delta),\;1/\eta_R\right\}>1$ and
$$\chi_X=\lim_{u\rightarrow1}\frac{\Pr\left(X_1>F^{-1}_{X_1}(u),X_2>F^{-1}_{X_2}(u) \right)}{1-u} \le \lim_{u\rightarrow1}\frac{(1-u)^{\min\left\{(\delta/(1-\delta),\;1/\eta_R\right\}}}{1-u}=0,$$
i.e. $(X_1,X_2)$ are AI, with $\eta_X\le\max\left\{(1-\delta)/\delta,\;\eta_R\right\}<1$.\vspace{0.3cm}\\
On the other hand, similarly to Case 1,
\begin{equation*}
\begin{split}
    \Pr\left(X_1>F^{-1}_{X_1}(u),X_2>F^{-1}_{X_2}(u) \right) \sim &\; \Pr\left(X_1>\left(\frac{\delta}{2\delta-1}\right)^\delta\left(\frac{1}{1-u}\right)^\delta,X_2>\left(\frac{\delta}{2\delta-1}\right)^\delta\left(\frac{1}{1-u}\right)^\delta \right)\\
    \ge&\; \Pr\left(\min\{R_1,R_2\} \; \min\{W_1,W_2\}^{(1-\delta)/\delta}>\frac{\delta}{(2\delta-1)(1-u)} \right)\\
    \sim&\; c(\delta)\, (1-u)^{\min\left\{1/\eta_R,\;\delta/[(1-\delta)\cdot\eta_W]\right\}},
\end{split}
\end{equation*}
with $c(\delta)>0$ for all $\delta>0.5$. If $(R_1,R_2)$ are AD, then $\eta_R=1$, so $\min\left\{1/\eta_R,\;\delta/[(1-\delta)\cdot\eta_W]\right\}=1$ and
$$\chi_X=\lim_{u\rightarrow1}\frac{\Pr\left(X_1>F^{-1}_{X_1}(u),X_2>F^{-1}_{X_2}(u) \right)}{1-u} \ge \lim_{u\rightarrow1}\frac{c(\delta)(1-u)}{1-u}>0,$$
i.e. $(X_1,X_2)$ are AD.

\medskip
\newpage
\subsection*{Case 3: \texorpdfstring{$\delta=0.5$}{delta=1/2} (second column of Table \ref{table:possible_models})}
\medskip

\noindent $(X_1,X_2)$ are AI, unless both $(W_1,W_2)$ and $(R_1,R_2)$ are AD, i.e.

\begin{itemize}
    \item[-]  $(X_1,X_2)\; \text{are AI, with}\; \eta_X<1 \qquad \text{if}\; (W_1,W_2), (R_1,R_2) \;\text{are AI, with}\; \eta_W<1,\eta_R<1$;
\item[-]  $(X_1,X_2)\; \text{are AI, with}\; \eta_X=1 \qquad \text{if}\; (W_1,W_2) \;\text{are AI, with}\; \eta_W<1,\; \text{and}\; (R_1,R_2)\;\text{are AD}$;
\item[-]  $(X_1,X_2)\; \text{are AI, with}\; \eta_X=1 \qquad \text{if}\; (W_1,W_2) \;\text{are AD and}\; (R_1,R_2)\;\text{are AI, with} \;\eta_R<1$;
\item[-]  $(X_1,X_2)\; \text{are AD} \qquad\qquad\qquad\quad\;\, \text{if}\; (W_1,W_2)\;\text{are AD and}\; (R_1,R_2) \;\text{are AD}$.
\end{itemize}

\bigskip

\underline{Proof}:
Since  $X_1=R_1^{1/2}W_1^{1/2}$ and $X_2=R_2^{1/2}W_2^{1/2}$ have the same marginal distribution, we can compute $\chi_X$ in \eqref{def:chi_X_12} as
$$\chi_X=\lim_{x\rightarrow\infty}\frac{\Pr\left(X_1>x,X_2>x\right)}{\Pr\left(X_1>x\right)}.$$
Note that, as reported in \eqref{marginal_distr_X}, the denominator is equal to
$$\Pr\left(X_1>x\right)=\;x^{-2}\left[2\log(x)+1\right].$$

\begin{description}
    \item If $(W_1,W_2)$ are AI with $\eta_W<1$ and $(R_1,R_2)$ are AD, the numerator is such that
\begin{equation*}
\begin{split}
    \Pr\left(X_1>x,X_2>x\right)=&\; \Pr\left(R_1^{1/2}W_1^{1/2}>x,R_2^{1/2}W_2^{1/2}>x\right)=\Pr\left(\min\left\{R_1W_1,R_2W_2\right\}>x^2\right)\\
    \leq&\; \Pr\left(R_1\,\min\{W_1,W_2\}>x^2\cap R_1>R_2\right)+\\
    &+\Pr\left(R_2\,\min\{W_1,W_2\}>x^2\cap R_1\le R_2\right)\\
    \leq&\; \Pr\left(R_1\,\min\{W_1,W_2\}>x^2\right)+ \Pr\left(R_2\,\min\{W_1,W_2\}>x^2\right)\\
    \sim&\; 2\, \E\left[\min\{W_1,W_2\}\right]\, \bar{F}_R(x^2) = 2\, \E\left[\min\{W_1,W_2\}\right]\, x^{-2}.
\end{split}
\end{equation*}
Therefore,
$$\chi_X=\lim_{x\rightarrow\infty}\frac{\Pr\left(X_1>x,X_2>x\right)}{\Pr\left(X_1>x\right)} \le \lim_{x\rightarrow\infty}\frac{2\, \E\left[\min\{W_1,W_2\}\right]\, x^{-2}}{x^{-2}\left[2\log(x)+1\right]}=0,$$
i.e. $(X_1,X_2)$ are AI, although $\eta_X=1$.
    \item If $(R_1,R_2)$ are AI with $\eta_R<1$ and $(W_1,W_2)$ are AD, the numerator is such that
\begin{equation*}
\begin{split}
    \Pr\left(X_1>x,X_2>x\right)=&\; \Pr\left(R_1^{1/2}W_1^{1/2}>x,R_2^{1/2}W_2^{1/2}>x\right)=\Pr\left(\min\left\{R_1W_1,R_2W_2\right\}>x^2\right)\\
    \leq&\; \Pr\left(W_1\,\min\{R_1,R_2\}>x^2\right)+ \Pr\left(W_2\,\min\{R_1,R_2\}>x^2\right)\\
    \sim&\; 2\, \E\left[\min\{R_1,R_2\}\right] \bar{F}_W(x^2) = 2\, \E\left[\min\{R_1,R_2\}\right] x^{-2}.
\end{split}
\end{equation*}
Therefore,
$$\chi_X=\lim_{x\rightarrow\infty}\frac{\Pr\left(X_1>x,X_2>x\right)}{\Pr\left(X_1>x\right)} \le \lim_{x\rightarrow\infty}\frac{2\, \E\left[\min\{R_1,R_2\}\right] x^{-2}}{x^{-2}\left[2\log(x)+1\right]}=0,$$
i.e. $(X_1,X_2)$ are AI, although $\eta_X=1$.
    \item If $(W_1,W_2)$ are AI with $\eta_W<1$ and $(R_1,R_2)$ are AI with $\eta_R<1$, the numerator is
$$\Pr\left(X_1>x,X_2>x\right) \sim c\; x^{-2\min\{1/\eta_W,\; 1/\eta_R\}},$$
so $\chi_X=0$ and $(X_1,X_2)$ are AI with $\eta_X=\max\{\eta_W,\eta_R\}<1$.
\item If both $(W_1,W_2)$ and $(R_1,R_2)$ are AD, the numerator is such that
\begin{equation*}
\begin{split}
  \Pr\left(X_1>x,X_2>x\right)=&\Pr\left(\min\left\{R_1^{1/2}W_1^{1/2},R_2^{1/2}W_2^{1/2}\right\}>x\right)\\
  \ge& \Pr\left(\min\{R_1,R_2\}^{1/2}\min\{W_1,W_2\}^{1/2}>x\right).
\end{split}
\end{equation*}
We use the result (1.5) from \cite{kasahara2018note}, proved by \cite{kifer2017tails}: 
if $Z_1, Z_2$ are independent random variables such that $$\Pr(Z_j>z)\sim c_j z^{-\alpha_j}(\log z)^{k_j},\quad j=1,2,$$ and $\alpha_1=\alpha_2$, then
$$\Pr(Z_1Z_2>z)\sim \alpha_1 B(k_1+1,k_2+1)c_1c_2z^{-\alpha_1}(\log z)^{k_1+k_2+1}.$$
Note that $$\Pr\left(\min\{R_1,R_2\}^{1/2}>x\right)\sim L_R(x^2)\, x^{-2}\quad\text{and} \quad \Pr\left(\min\{W_1,W_2\}^{1/2}>x\right)\sim L_W(x^2)\, x^{-2},$$
$$\text{with}\quad L_R(x^2)\rightarrow \chi_R>0\quad\text{and} \quad L_W(x^2)\rightarrow \chi_W>0,\; \text{as}\; x\rightarrow\infty,$$
i.e. there exist a positive constant $\varepsilon<\min\{\chi_R,\chi_W\}$ and $x_\varepsilon$ such that, for all $x>x_\varepsilon$,
$$L_R(x^2)\ge \chi_R-\varepsilon \quad\text{and} \quad L_W(x^2)\ge \chi_W-\varepsilon.$$
Then, for large $x$,
$$\Pr\left(\min\{R_1,R_2\}^{1/2}\, \min\{W_1,W_2\}^{1/2}>x\right)\ge c\, x^{-2}\log x,$$
with $c=(\chi_R-\varepsilon)(\chi_W-\varepsilon)$, and
$$\chi_X=\lim_{x\rightarrow\infty}\frac{\Pr\left(X_1>x,X_2>x\right)}{\Pr\left(X_1>x\right)} \ge c>0,$$
i.e. $(X_1,X_2)$ are AD.
\end{description}

\section{Proof on dependence properties (pairs in space)}\label{section:appendix_space}

This appendix contains the proofs regarding the dependence properties of model \eqref{model_X_RW} that are stated in Section \ref{section:dependence_properties} for pairs of observations in different locations and contemporaneous times.
Let $(X_1,X_2)\,=(R_1^{\delta}W_1^{1-\delta},R_2^{\delta}W_2^{1-\delta})$, with $X_{i}:=X(s_i,t_i)$, $R_i:=R(t_i)$ and $W_{i}:=W(s_i,t_i)$, for $i=1,2$, such that $s_1\ne s_2$ and $t_1=t_2$, be a generic spatial pair from model \eqref{model_X_RW}. Since $R_1=R_2$, we refer to them as $R$ in the following. These proofs are based on some modifications of those in \ref{section:appendix_time_spacetime}.

\bigskip
\subsection*{Case 1: \texorpdfstring{$\delta<0.5$}{delta<1/2} (third column of Table \ref{table:possible_models})}
\medskip

\noindent$(X_1,X_2)$ belong to the same extremal dependence class as $(W_1,W_2)$, i.e.,

\begin{itemize}
    \item[-]  $(X_1,X_2)\; \text{are AI} \qquad\qquad\qquad\quad\;\;\; \text{if}\; (W_1,W_2) \;\text{are AI, with}\; \eta_W<1$;
    \item[-]  $(X_1,X_2)\; \text{are AD} \qquad\qquad\qquad\quad\:\: \text{if}\; (W_1,W_2) \;\text{are AD}$.
\end{itemize}

\bigskip

\underline{Proof}:
Similarly to Case 1 in \ref{section:appendix_time_spacetime}, as $u\rightarrow1$, the numerator of $\chi_X$ is
\begin{equation*}
\begin{split}
    \Pr\left(X_1>F^{-1}_{X_1}(u),X_2>F^{-1}_{X_2}(u) \right) <&\; \Pr\left(\min\left\{R^{\delta/(1-\delta)} W_1,R^{\delta/(1-\delta)} W_2\right\}>\frac{1}{1-u} \right)\\
    =&\; \Pr\left(R^{\delta/(1-\delta)}\min\{W_1,W_2\}>\frac{1}{1-u} \right)\\
    \sim& \; (1-u)^{\min\left\{(1-\delta)/\delta,\;1/\eta_W\right\}}.
\end{split}
\end{equation*}
The last line derives from Breiman's Lemma, since, as $u\rightarrow1$, $\Bar{F}_{\min\{W_1,W_2\}}(x)\sim x^{-1/\eta_W}$, and $\Bar{F}_{R^{\delta/(1-\delta)}}(x)\sim x^{-(1-\delta)/\delta}$.\vspace{0.3cm}\\
 If $(W_1,W_2)$ are AI with $\eta_W<1$, $\min\left\{(1-\delta)/\delta,\;1/\eta_W\right\}>1$ and
$$\chi_X=\lim_{u\rightarrow1}\frac{\Pr\left(X_1>F^{-1}_{X_1}(u),X_2>F^{-1}_{X_2}(u) \right)}{1-u} \le \lim_{u\rightarrow1}\frac{(1-u)^{\min\left\{(1-\delta)/\delta,1/\eta_W\right\}}}{1-u}=0,$$
i.e. $(X_1,X_2)$ are AI, with $\eta_X\le \max\left\{\delta/(1-\delta),\;\eta_W\right\}<1$.\vspace{0.3cm}\\
On the other hand,
\begin{equation*}
\begin{split}
    \Pr\left(X_1>F^{-1}_{X_1}(u),X_2>F^{-1}_{X_2}(u) \right) \sim &\; \Pr\left(R^{\delta/(1-\delta)}\;\min\{W_1,W_2\}>\frac{1-\delta}{(1-2\delta)(1-u)} \right)\\
    \sim&\; c(\delta)\, (1-u)^{\min\left\{(1-\delta)/\delta,\; 1/\eta_W\right\}},
\end{split}
\end{equation*}
with $c(\delta)>0$ for all $\delta<0.5$.
\vspace{0.3cm}\\
If $(W_1,W_2)$ are AD, then $\eta_W=1$, so $\min\left\{(1-\delta)/\delta,\; 1/\eta_W\right\}=1$ and
$$\chi_X=\lim_{u\rightarrow1}\frac{\Pr\left(X_1>F^{-1}_{X_1}(u),X_2>F^{-1}_{X_2}(u) \right)}{1-u} \ge \lim_{u\rightarrow1}\frac{c(\delta)(1-u)}{1-u}>0,$$
i.e. $(X_1,X_2)$ are AD.

\bigskip
\subsection*{Case 2: \texorpdfstring{$\delta>0.5$}{delta>1/2} (first column of Table \ref{table:possible_models})}
\bigskip

\noindent$(X_1,X_2)$ \text{are AD}.

\bigskip
\underline{Proof}: 
Similarly to Case 2 in \ref{section:appendix_time_spacetime}, as $u\rightarrow1$, the numerator of $\chi_X$ is
\begin{equation*}
\begin{split}
    \Pr\left(X_1>F^{-1}_{X_1}(u),X_2>F^{-1}_{X_2}(u) \right) \ge&\; \Pr\left(R \; \min\{W_1,W_2\}^{(1-\delta)/\delta}>\frac{\delta}{(2\delta-1)(1-u)} \right)\\
    \sim&\; c(\delta)\, (1-u)^{\min\left\{1,\;\delta/[(1-\delta)\cdot\eta_W]\right\}}
\end{split}
\end{equation*}
with $c(\delta)>0$  for all $\delta>0.5$. Since $\min\left\{1,\;\delta/[(1-\delta)\cdot\eta_W]\right\}=1$,
$$\chi_X=\lim_{u\rightarrow1}\frac{\Pr\left(X_1>F^{-1}_{X_1}(u),X_2>F^{-1}_{X_2}(u) \right)}{1-u} \ge \lim_{u\rightarrow1}\frac{c(\delta)(1-u)}{1-u}>0,$$
i.e. $(X_1,X_2)$ are AD.

\medskip
\subsection*{Case 3: \texorpdfstring{$\delta=0.5$}{delta=1/2} (second column of Table \ref{table:possible_models})}
\medskip

\noindent$(X_1,X_2)$ belong to the same extremal dependence class as $(W_1,W_2)$, i.e.,
\begin{itemize}
    \item[-]  $(X_1,X_2)\; \text{are AI, with}\; \eta_X=1  \qquad\qquad \text{if}\; (W_1,W_2) \;\text{are AI, with}\; \eta_W<1$;
    \item[-]  $(X_1,X_2)\; \text{are AD} \qquad\qquad\qquad\qquad\quad\:\: \text{if}\; (W_1,W_2) \;\text{are AD}$.
\end{itemize}

\bigskip

\underline{Proof}:
As in Case 3 of \ref{section:appendix_time_spacetime}, the denominator of $\chi_X$ is equal to
$$\Pr\left(X_1>x\right)=\;x^{-2}\left[2\log(x)+1\right].$$
\noindent If $(W_1,W_2)$ are AI with $\eta_W<1$, the numerator is such that
\begin{equation*}
\begin{split}
    \Pr\left(X_1>x,X_2>x\right)=&\; \Pr\left(R^{1/2}W_1^{1/2}>x,R^{1/2}W_2^{1/2}>x\right)=\Pr\left(R\min\left\{W_1,W_2\right\}>x^2\right)\\
    \sim&\; \E\left[\min\{W_1,W_2\}\right]\, \bar{F}_R(x^2) = \E\left[\min\{W_1,W_2\}\right]\, x^{-2}.
\end{split}
\end{equation*}
Therefore,
$$\chi_X=\lim_{x\rightarrow\infty}\frac{\Pr\left(X_1>x,X_2>x\right)}{\Pr\left(X_1>x\right)} = \lim_{x\rightarrow\infty}\frac{\E\left[\min\{W_1,W_2\}\right]\, x^{-2}}{x^{-2}\left[2\log(x)+1\right]}=0,$$
i.e. $(X_1,X_2)$ are AI, although $\eta_X=1$.\vspace{0.3cm}

\noindent If $(W_1,W_2)$ are AD, the numerator is such that
\begin{equation*}
\begin{split}
  \Pr\left(X_1>x,X_2>x\right)=&\Pr\left(R^{1/2}\min\left\{W_1^{1/2},W_2^{1/2}\right\}>x\right)\\
  \ge&\; c\, x^{-2}\log x,
\end{split}
\end{equation*}
with $c=\chi_W-\varepsilon$, for small $\varepsilon>0$. The last inequality is derived using the same results as in the last point of Case 3 in \ref{section:appendix_time_spacetime}, since $\Pr\left(R^{1/2}>x\right)=x^{-2}$. Then,
$$\chi_X=\lim_{x\rightarrow\infty}\frac{\Pr\left(X_1>x,X_2>x\right)}{\Pr\left(X_1>x\right)} \ge c>0,$$
i.e. $(X_1,X_2)$ are AD.








\renewcommand{\theequation}{S\arabic{equation}}
\renewcommand{\thefigure}{S\arabic{figure}}
\renewcommand{\thetable}{S\arabic{table}}
\renewcommand{\thesection}{S\arabic{section}}




\MakeTitle{Supplementary material}{}{}

\section{Simulation study}

\begin{figure}[h!]
\begin{subfigure}[h]{0.5\linewidth}
\includegraphics[width=\linewidth]{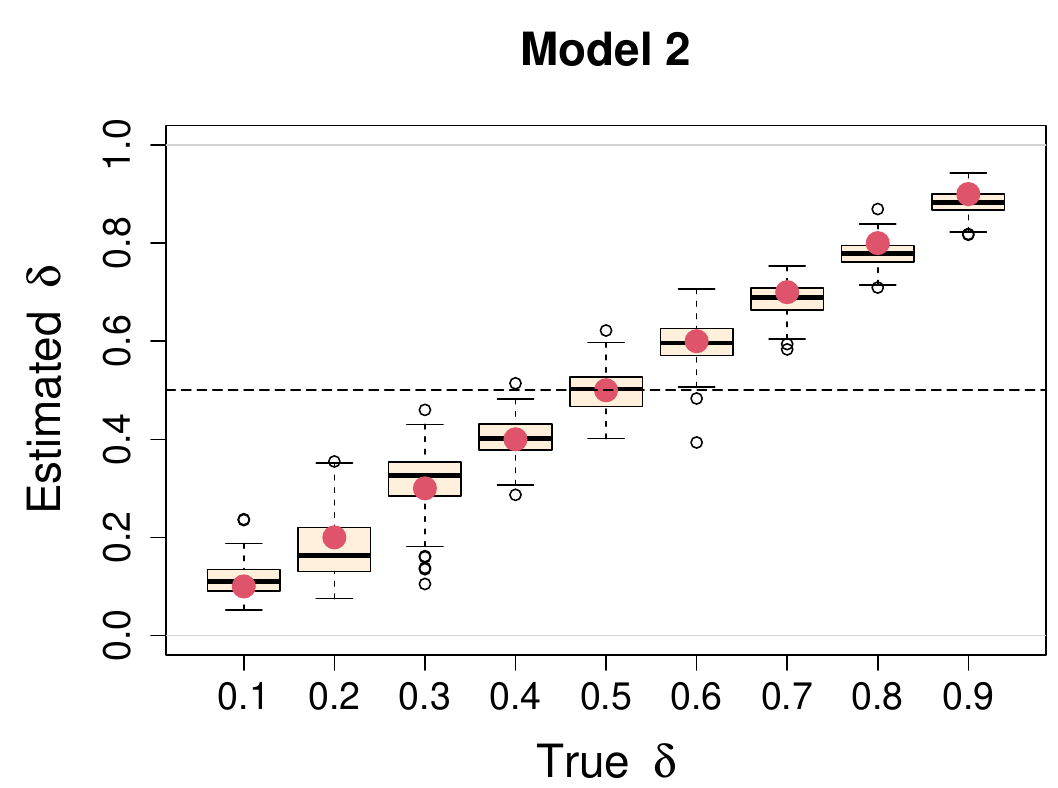}
\end{subfigure}
\hfill
\begin{subfigure}[h!]{0.5\linewidth}
\includegraphics[width=\linewidth]{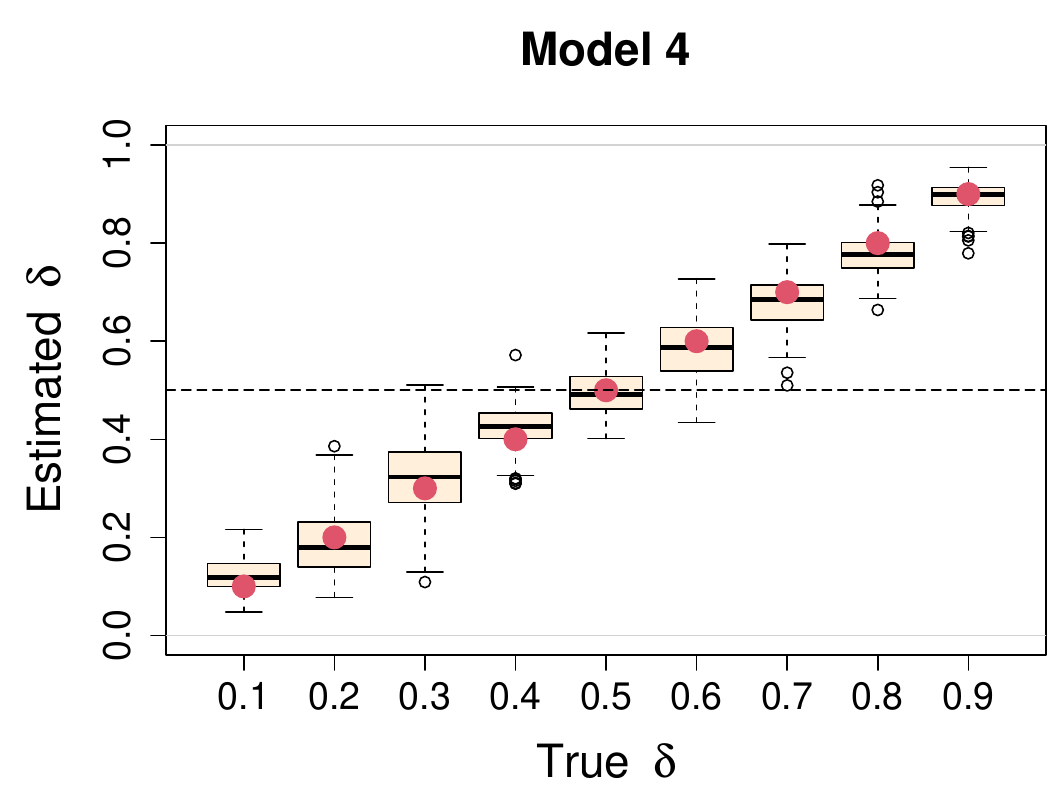}
\end{subfigure}%
\caption{True (red points) and estimated (boxplots) values of $\delta$ on 200 independently simulated datasets, for Model 2 (left) and Model 4 (right).}
\label{fig:sim_study_delta_2_4}
\end{figure}

Figure \ref{fig:sim_study_delta_2_4} shows the estimated values for $\delta$ in $\{0.1,0.2,\dots,0.9\}$ for Model 2 and Model 4, to be compared with Figure \ref{fig:sim_study_delta} in the main text (Model 1 and Model 3). Among the four models, the one on which the estimation seems to have the worst performance is Model 4, in which all the components of the mixture are AD and $\delta$ does not discriminate between extremal dependence classes (see Table \ref{table:possible_models} in the main text).

\begin{figure}[h!]
\includegraphics[width=\linewidth]{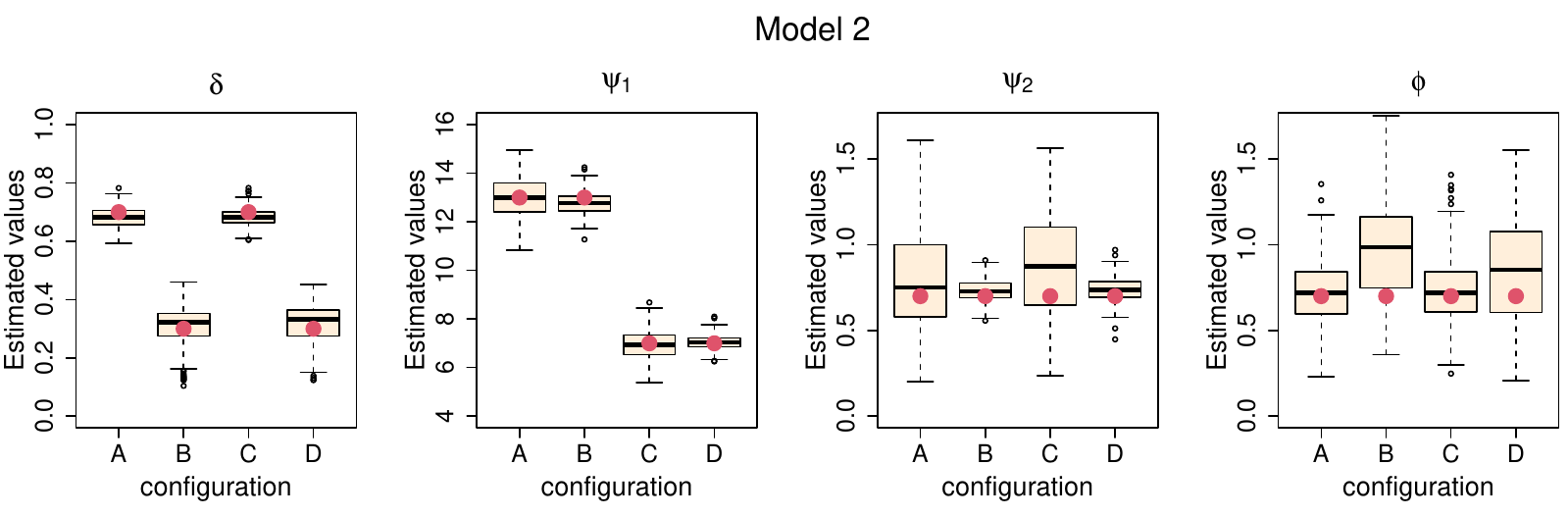}
\includegraphics[width=\linewidth]{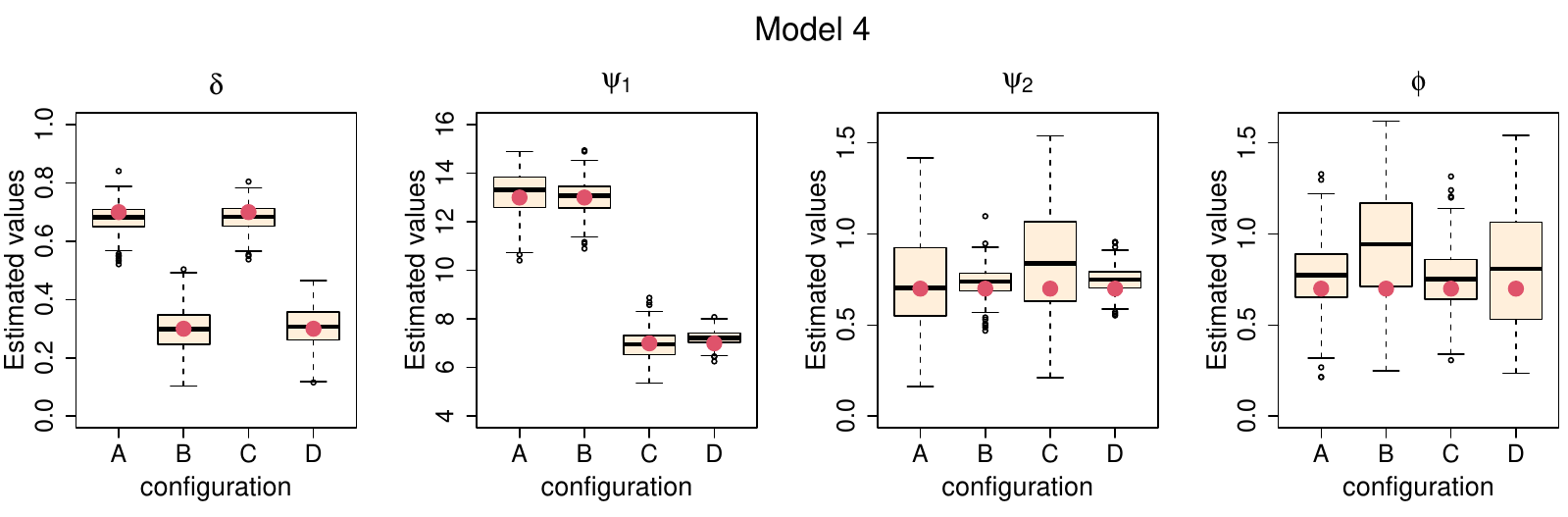}
\caption{Boxplots of estimated values for $\delta=0.7$ (A,C) or $\delta=0.3$ (B,D), $\psi_1=13$ (A,B) or $\psi_1=7$ (C,D), $\psi_2=0.7$ and $\phi=0.7$ (red points) on 400 simulated datasets from Model 2 and Model 4.}
\label{fig:sim_study_psi_phi_2_4}
\end{figure}

\newpage
Figure \ref{fig:sim_study_psi_phi_2_4} shows the boxplots relative to the simulation study on $\psi_1$, $\psi_2$ and $\phi$ for Model 2 and Model 4, to be compared with Figure \ref{fig:sim_study_psi_phi} in the main text (Model 1 and Model 3).

\section{Results and diagnostics}

This section presents some results about the real data application with a correlation function different from that presented in section \ref{section:application_model}. In particular, the correlation function of the Gaussian process on which $W(s,t)$ is built is here $\rho_{W^*}(h,k;\psi)= [1+(\|h\|/\psi_1)^2]^{-1} \times \exp\{-(k/\psi_2)^2\}$, for spatial lag $h$ and temporal lag $k$.

\begin{figure}[ht!]
\begin{subfigure}{0.329\linewidth}
\includegraphics[width=\linewidth]{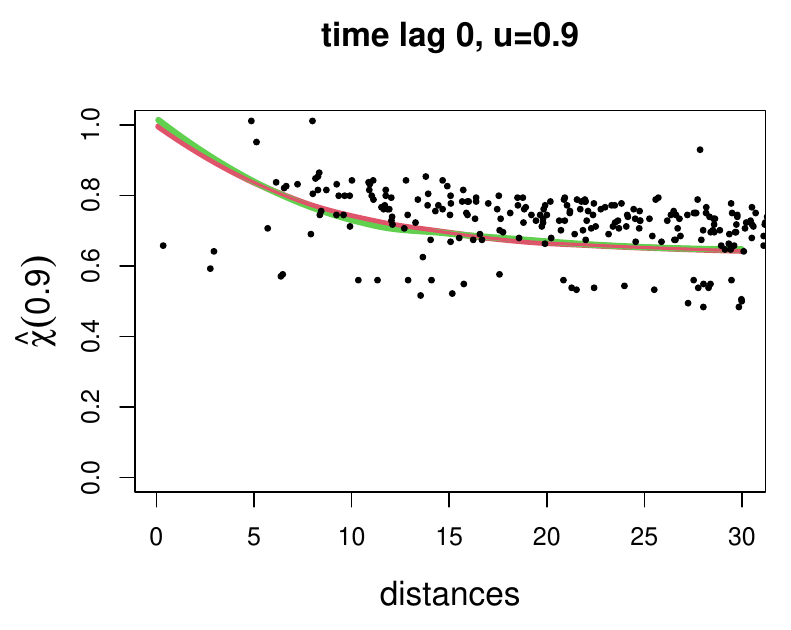}
\end{subfigure}
\begin{subfigure}{0.329\linewidth}
\includegraphics[width=\linewidth]{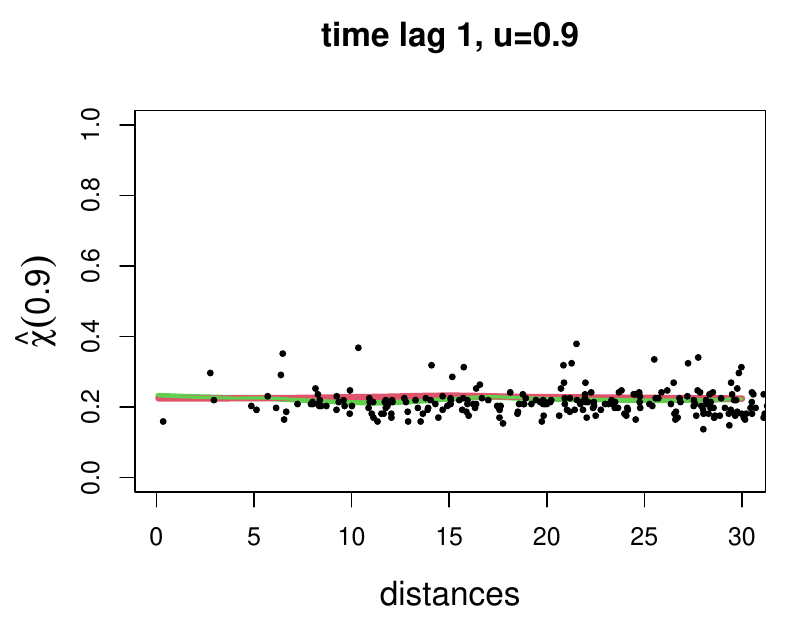}
\end{subfigure}
\begin{subfigure}{0.329\linewidth}
\includegraphics[width=\linewidth]{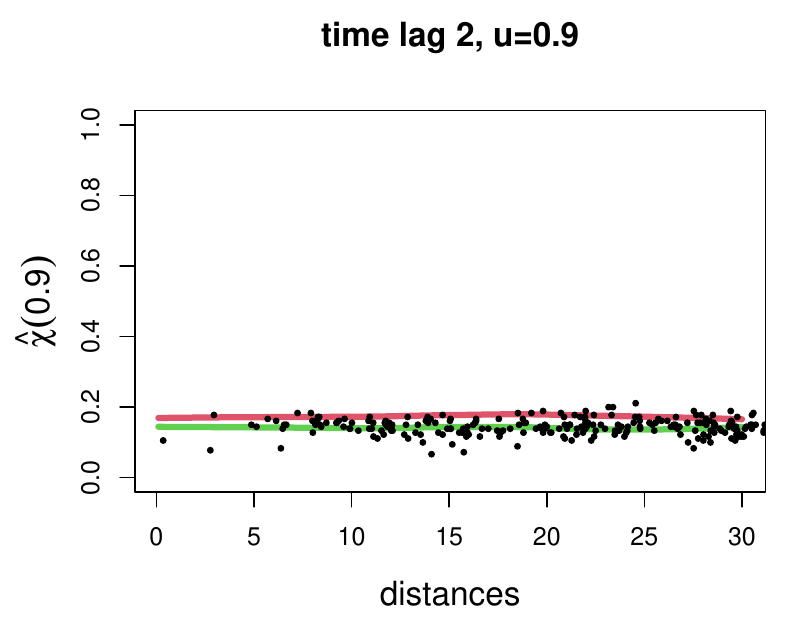}
\end{subfigure}

\begin{subfigure}{0.329\linewidth}
\includegraphics[width=\linewidth]{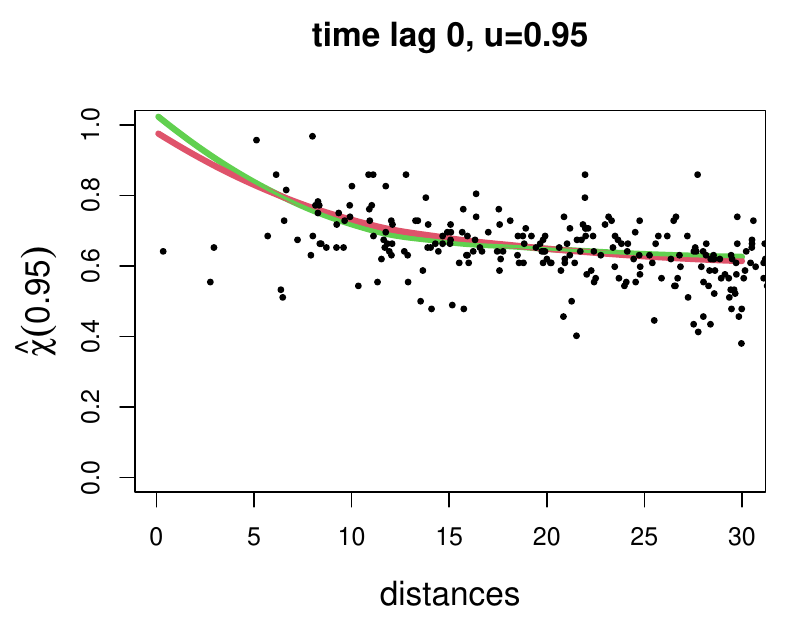}
\end{subfigure}
\begin{subfigure}{0.329\linewidth}
\includegraphics[width=\linewidth]{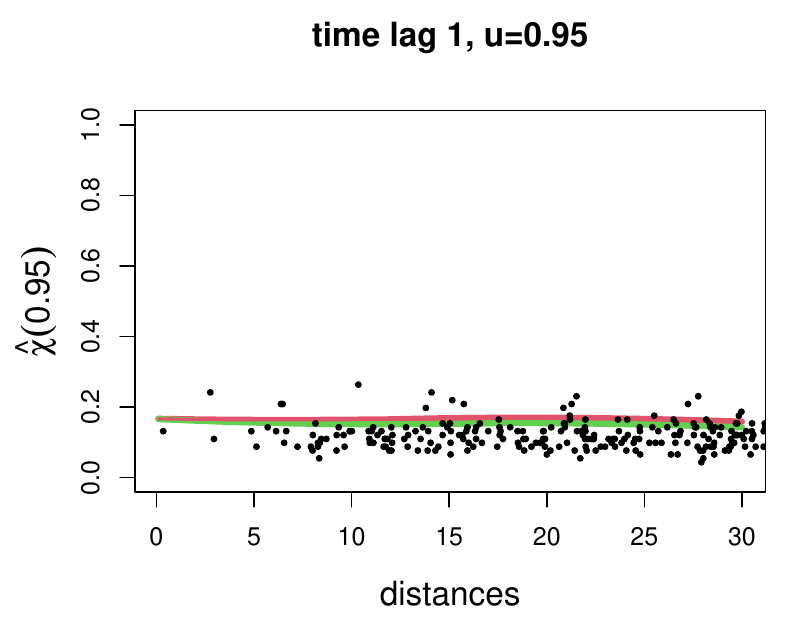}
\end{subfigure}
\begin{subfigure}{0.329\linewidth}
\includegraphics[width=\linewidth]{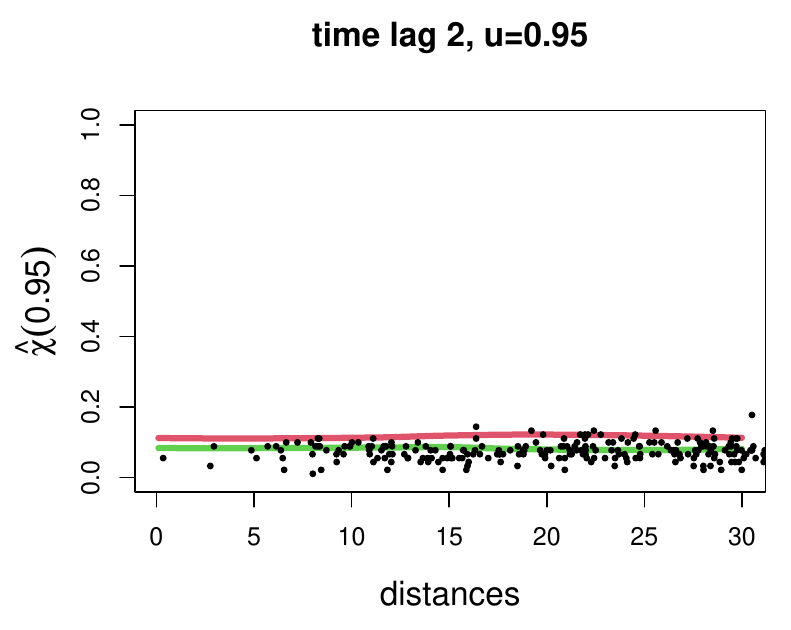}
\end{subfigure}

\begin{subfigure}{0.329\linewidth}
\includegraphics[width=\linewidth]{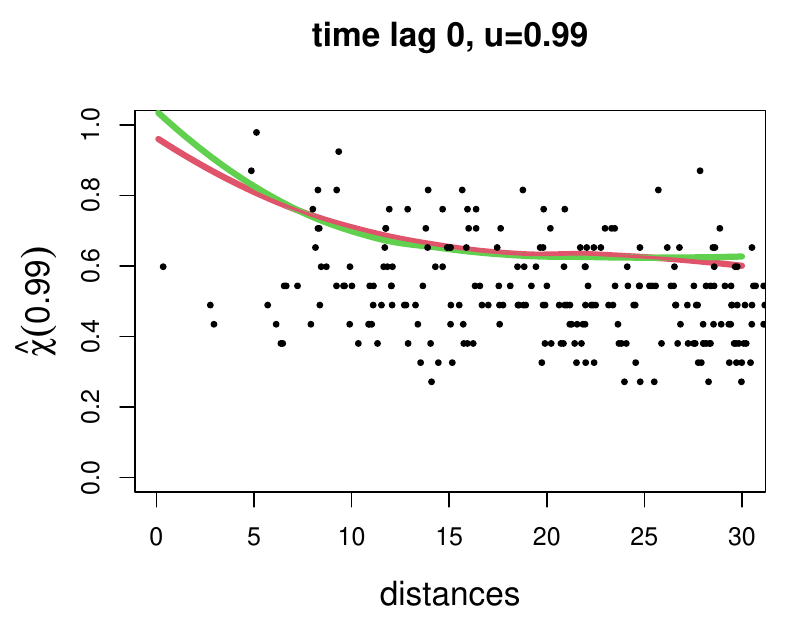}
\end{subfigure}
\begin{subfigure}{0.329\linewidth}
\includegraphics[width=\linewidth]{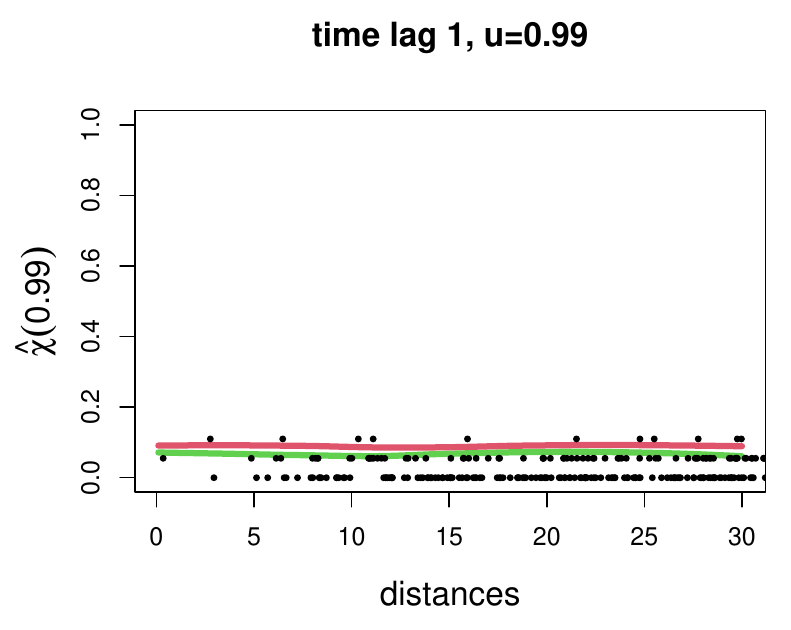}
\end{subfigure}
\begin{subfigure}{0.329\linewidth}
\includegraphics[width=\linewidth]{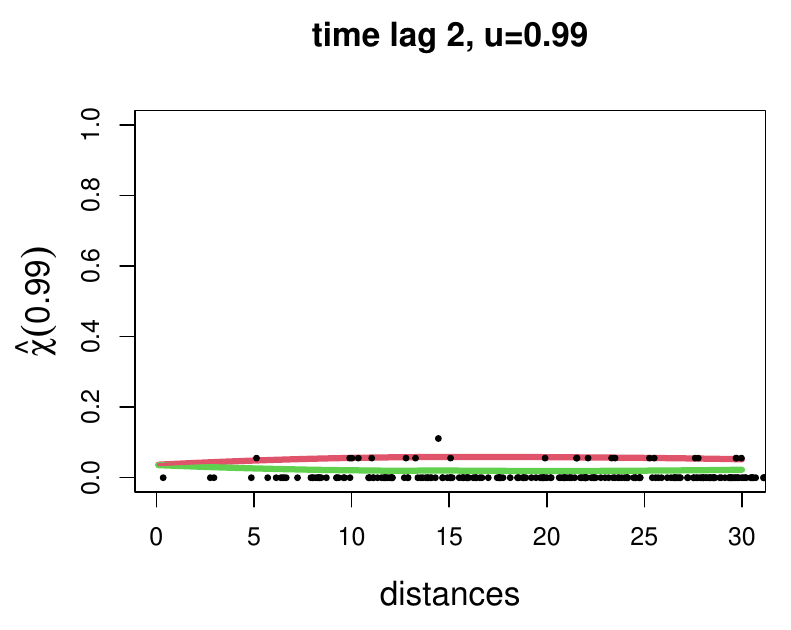}
\end{subfigure}
\caption{Values of  ${\chi(u)}$  for different spatial distances and temporal lags, derived from the estimated Model 1 (red lines), from the Model 3 (green lines) and the empirical ones for each pair of stations (dots).}
\label{fig:diagn_mod_1_3_ExpGaussian}
\end{figure}

Figure \ref{fig:diagn_mod_1_3_ExpGaussian}  shows the empirical pairwise coefficient $\hat{\chi}_{1,2}(u)$, for pairs $(Y_1,Y_2)$ at different distances in space and time and for $u\in\{0.9,0.95,0.99\}$. Compared with Figure \ref{fig:diagn_mod_1_3} in the main text, it does not present substantial differences.

Table \ref{table:RMSE_corGau} displays the $\text{RMSE}(k,u)$ for Model 1 and Model 3, to be compared with those of Table \ref{table:RMSE} in the main text. The values indicate a slight preference for the correlation function presented in the text, compared to the one presented here.

\begin{table}[h!]
\begin{center}
\begin{tabular}{| c | c | c | c | c | c | c |} 
 \hline
  \multirow{2}{*}{Model} & \multicolumn{2}{c|}{$k=0$} & \multicolumn{2}{c|}{$k=1$} & \multicolumn{2}{c|}{$k=2$} \\
  \cline{2-7} 
   & $u=0.9$ & $u=0.95$ & $u=0.9$ & $u=0.95$ & $u=0.9$ & $u=0.95$ \\ [0.5ex] 
 \hline
 1 & 0.115 & 0.145 & 0.026 & 0.028 & 0.027 & 0.029 \\
 \hline
 3 & 0.108 & 0.113 & 0.029 & 0.031 & 0.028 & 0.030  \\
 \hline
\end{tabular}
\caption{$\text{RMSE}(k,u)$ between empirical estimates of $\chi^*_{s_i;k}(u)$ and estimates from Model 1 and Model 3.}
\label{table:RMSE_corGau}
\end{center}
\end{table}

\clearpage
\section{Marginal parameter estimation}\label{section:suppl_marginal}
\begin{figure}[h!]
\begin{subfigure}[h]{0.5\linewidth}
\includegraphics[width=\linewidth]{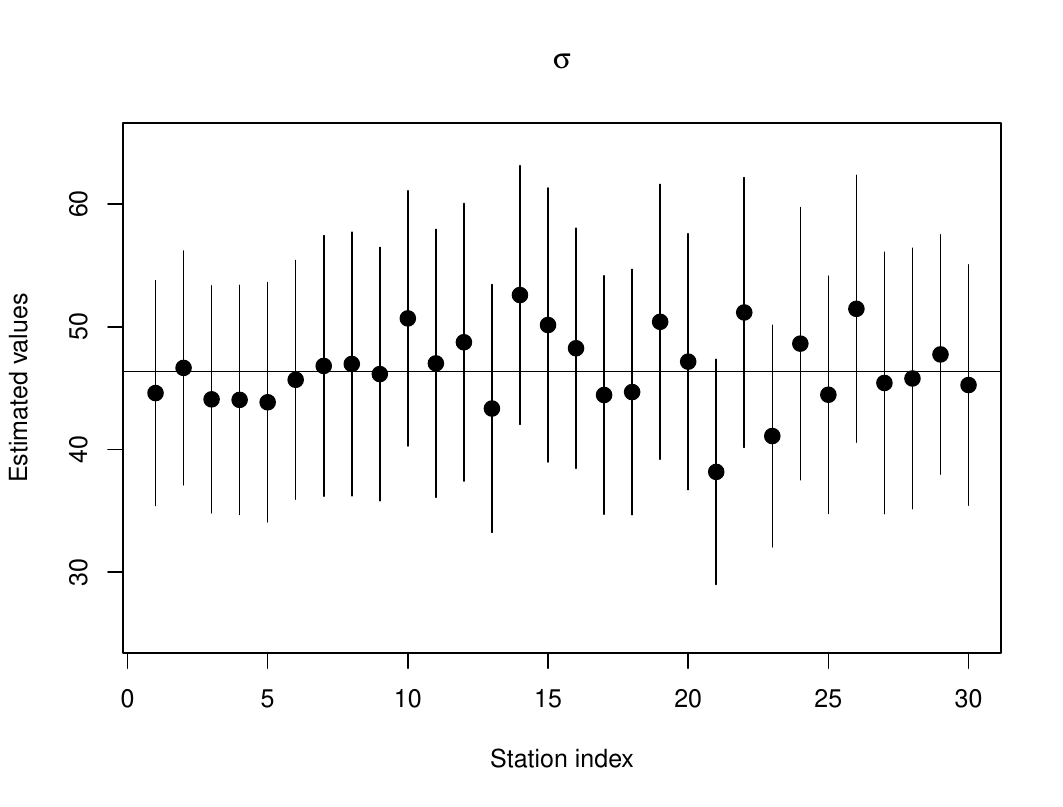}
\end{subfigure}
\hfill
\begin{subfigure}[h!]{0.5\linewidth}
\includegraphics[width=\linewidth]{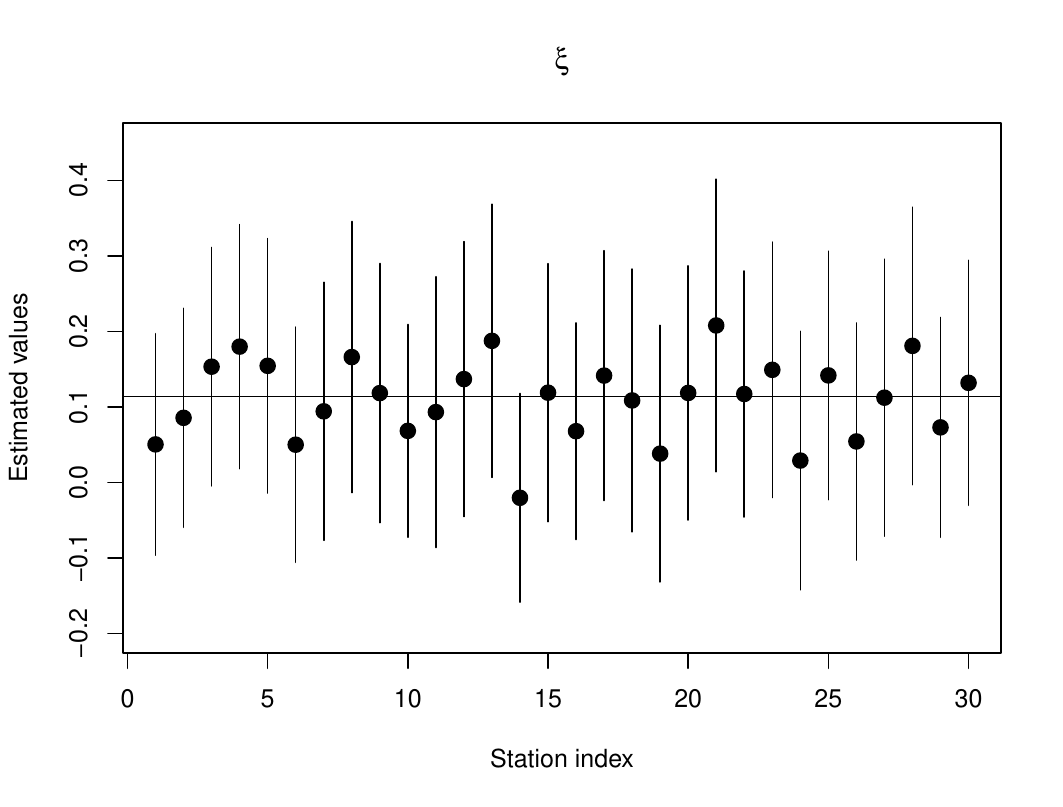}
\end{subfigure}%
\caption{Site-wise estimates (points) and $95\%$ confidence intervals (vertical segments) for $\sigma$ and $\xi$. The horizontal lines denote the values estimated on all stations together.}
\label{fig:est_sigma_xi_sitewise}
\end{figure}

Figure \ref{fig:est_sigma_xi_sitewise} shows site-specific estimates and $95\%$ confidence intervals for the marginal parameter $\sigma$ and $\xi$. The horizontal lines represent the estimates of the two parameters obtained by fitting the GPD model on all the threshold exceedances together. The horizontal lines intersect all of the site-wise confidence
intervals, supporting the hypothesis of constant $\sigma$ and $\xi$ for the North Brabant data.

\clearpage
\section{Alternative models}

The next Table \ref{table:possible_models_5678} is the equivalent of Table \ref{table:possible_models} for the alternative models originating from equation \eqref{eq:model_5678} in the main text, which we recall here:
$$X(s,t)=R(s)^\delta\, W(s,t)^{1-\delta},\quad \delta\in[0,1].$$
To prove these results, follow the appendices in the article, carefully inverting the roles of space and time.
Results referring to space and space-time can be proven by following  \ref{section:appendix_time_spacetime}, while results referring to pairs in time can be proven by following  \ref{section:appendix_space}.

\begin{table}[h!]
	\begin{center}
		\begin{tabular}{| c | c | c | c | c | c |} 
			\hline
			Model & $R(s)$ & $ W(s,t)$ & $\delta>0.5$ & $\delta=0.5$ & $\delta<0.5$ \\ [0.5ex] 
			\hline\hline
			5 & AI & AD & \makecell[l]{AI in space\\ AD in time\\AI in space-time} & \makecell[l]{AI in space\\ AD in time\\AI in space-time} & \makecell[l]{AD in space\\ AD in time\\AD in space-time} \\ 
			\hline
			6 & AD & AI & \makecell[l]{AD in space\\ AD in time\\AD in space-time} & \makecell[l]{AI in space\\ AI in time\\AI in space-time} & \makecell[l]{AI in space\\ AI in time\\AI in space-time} \\ 
			\hline
			7 & AI & AI & \makecell[l]{AI in space\\ AD in time\\AI in space-time} & \makecell[l]{AI in space\\ AI in time\\AI in space-time} & \makecell[l]{AI in space\\ AI in time\\AI in space-time} \\ 
			\hline
			8 & AD & AD & \makecell[l]{AD in space\\ AD in time\\AD in space-time} & \makecell[l]{AD in space\\ AD in time\\AD in space-time} & \makecell[l]{AD in space\\ AD in time\\AD in space-time} \\ 
			\hline
		\end{tabular}
		\caption{The four possible combinations for  $R(s)$ and $W(s,t)$ in \eqref{eq:model_5678} and the resulting extremal dependence for pairs $[X(s_1,t_1),X(s_2,t_2)]$ in space, in time and in space-time, for different values of the parameter $\delta$.}
		\label{table:possible_models_5678}
	\end{center}
\end{table}

\end{document}